\documentclass[pra,reprint,doublecolumn,superscriptaddress, nobalancelastpage,dvipsnames]{revtex4-2}
\usepackage[dvipsnames]{xcolor}

\usepackage{tikz}
\usepackage{tikz-feynman}
\usepackage{pgfplots}
\usepackage[sort&compress]{natbib}
\usepackage{verbatim}
\usepackage{float}
\usepackage{sidecap}
\sidecaptionvpos{figure}{c}
\usepackage{lipsum}
\usepackage{capt-of}
\usepackage{setspace}
\usepackage{dcolumn}
\usepackage{bm}
\usepackage{amssymb}
\usepackage{amsmath}
\usepackage{mathtools}
\usepackage{booktabs}
\usepackage{physics}
\usepackage{multirow}
\usepackage[a4paper]{geometry}
\newgeometry{top=1.25cm,right=1cm,left=1.25cm,bottom=1.25cm}
\usepackage[colorlinks=true, urlcolor=blue, pdfborder={0 0 0}]{hyperref}
\hypersetup{linkcolor=blue,citecolor=BrickRed}

\newcommand{\x}{\boldsymbol{x}}

\draft%
\begin{document}

%\title{\Large{\textbf{B Meson Flavour Tagging via Quantum Machine Learning}}}
%\title{\Large{\textbf{Boosted Ensembles of Qubit and Continuous Variable Quantum \\Support Vector Machines for B Meson Flavour Tagging }}}
\title{\Large{\normalfont{Boosted Ensembles of Qubit and Continuous Variable Quantum \\Support Vector Machines for B Meson Flavour Tagging }}}

\author{Maxwell T. West} \email{westm2@student.unimelb.edu.au}  \affiliation{School of Physics, The University of Melbourne, Parkville, 3010, VIC, Australia}
\author{Martin Sevior} \affiliation{School of Physics, The University of Melbourne, Parkville, 3010, VIC, Australia}
\author{Muhammad Usman} \email{musman@unimelb.edu.au}  \affiliation{School of Physics, The University of Melbourne, Parkville, 3010, VIC, Australia}
\affiliation{Data61, CSIRO, Clayton, 3168, VIC, Australia}

\maketitle%

\onecolumngrid%

%\textcolor{black}{\normalsize{\textbf{The increasingly widespread deployment of machine learning (ML) algorithms for scientific, technological,
%industrial and military applications has seen the reliability and robustness of ML come under heavy scrutiny in recent years.  }}}
\noindent%
\textcolor{black}{\normalsize{\textbf{The recent physical realisation of quantum computers with hundreds of noisy qubits has given
birth to an intense search for useful applications of their unique capabilities. One area that has received
particular attention is quantum machine learning (QML), the study of machine learning algorithms running
natively on quantum computers.
In this work we develop and apply QML methods to B meson flavour tagging, an important component of
experiments which probe CP violation in order to better
understand the matter-antimatter asymmetry of the universe. We simulate boosted ensembles
of quantum support vector machines (QSVMs) based on both conventional qubit-based and continuous
variable architectures, attaining effective tagging efficiencies of 28.0\% and 29.2\% respectively, comparable
with the leading published result of 30.0\% using classical machine learning algorithms. The ensemble nature
of our classifier is of particular importance, doubling the effective tagging efficiency of a single QSVM, which
we find to be highly prone to overfitting. These results are obtained despite the constraint of working
with QSVM architectures that are classically simulable, and we find evidence that QSVMs
beyond the simulable regime may be able to realise even higher performance, when sufficiently 
powerful quantum hardware is developed to execute them.
  }}}
\\ \\ \\
%While early work has suggested that QML methods will also be vulnerable to the existence of  adversarial examples, we
%argue that such examples may be difficult to construct classically, due to the potential reliance of a QML agent on
%features of the data which cannot be efficiently discovered classically, and produce early empirical evidence of such
%robustness. This resistance to classical adversarial attacks may confer a significant practical advantage to early adopters of powerful quantum computers.}}}
\twocolumngrid%
\noindent%
\section{Introduction}
Machine learning (ML) has become an important tool in modern physics due to its ability
to find the often faint signals indicative of interesting or rare processes~\cite{petabyte_highway}.
This is particularly true in particle physics,
where efforts to resolve the few lingering inconsistencies between the Standard Model and experimental results 
typically involve the 
intense analysis of small deviations from theoretical expectations~\cite{capdevila2018patterns,aaij2020measurement,aebischer2020b}.
Indeed, ML has been widely and successfully employed across experimental particle physics in recent 
years~\cite{flavourtagging,mlparticlephysics,andrews2020end,fbdt,bfactories,hepmllivingreview,LARKOSKI20201,guest2018deep,bourilkov2019machine,karagiorgi2022machine,baldi2016parameterized}.
At the same time, the emergence of programmable quantum computers has led to intense interest in the newborn field of quantum machine learning
~\cite{biamonte2017quantum,beer2020training, havlivcek2019supervised, romero2017quantum, dallaire2018quantum,killoran2019continuous,schuld2019quantum,qcnn,schuld2021supervised,nguyen2022theory,schatzki2022theoretical,west2023benchmarking,incudini2023resource,west2023towards}
(QML), which under some circumstances may  
offer the potential for quantum advantage in ML tasks even on noisy intermediate-scale quantum (NISQ) computers~\cite{huang2022quantum}.
Following the successful application of classical ML algorithms, the potential for QML to yield new benefits in high energy physics 
has already begun to be explored, with promising, if preliminary, results reported to date~\cite{blance2021quantum,guan2021quantum,terashi2021event,heredge2021quantum,blance2021unsupervised,mott2017solving,wu2021application,rousselot2023generative,tuysuz2021hybrid,wu2022challenges,chen2022quantum}.\\

\begin{figure*}[ht]
\begin{center}
\begin{tikzpicture}
\node[inner sep=0pt] (carch) at (-1.85,.7)
    {\includegraphics[width=10.15cm]{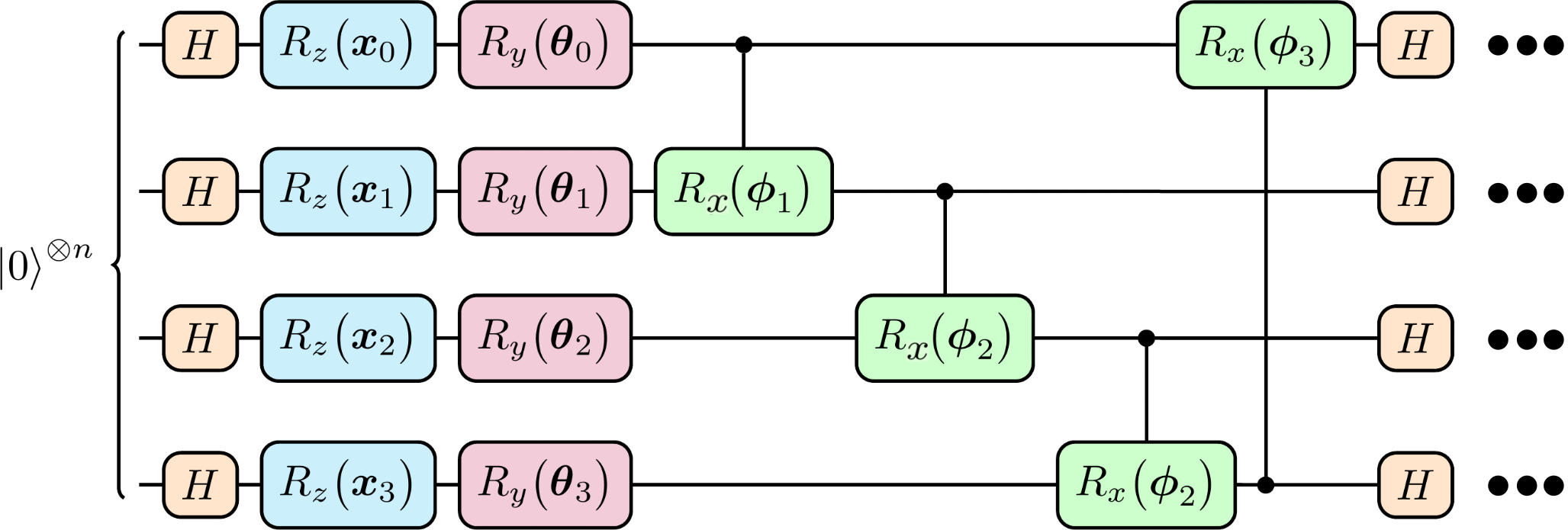}};
\node[inner sep=0pt] (varch) at (-1.85,-4.5)
    {\includegraphics[width=10cm]{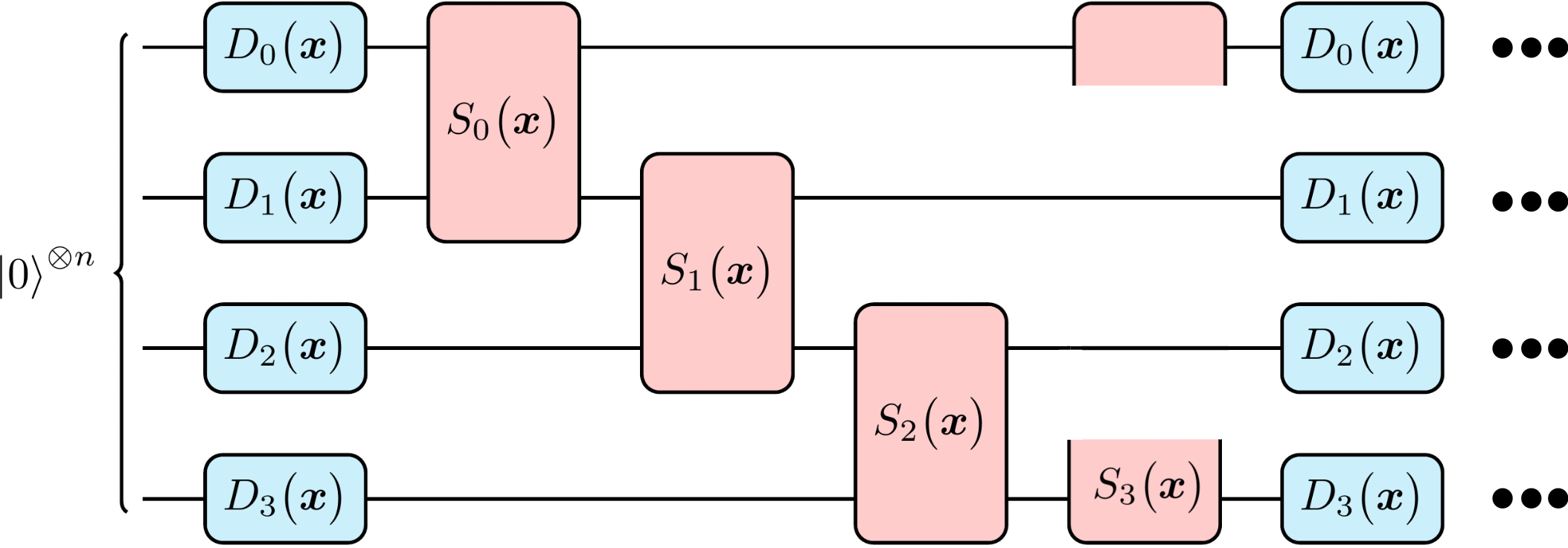}};
\node[inner sep=0pt] (embed) at (6.1,-4.35)
    {\includegraphics[width=3.5cm]{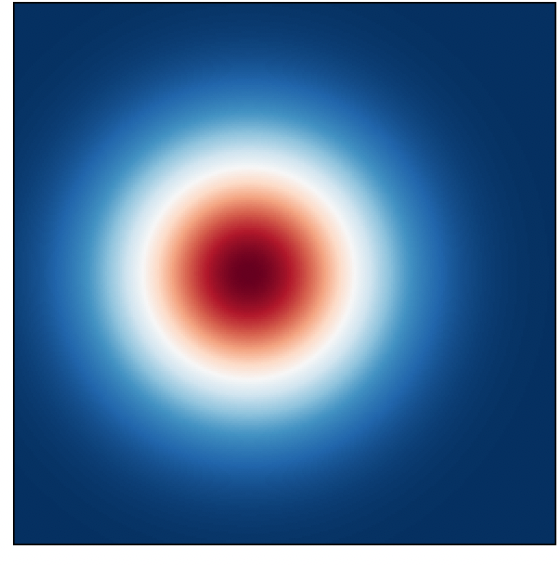}};
\node[inner sep=0pt] (embed) at (10.5,-4.3)
    {\includegraphics[width=3.5cm]{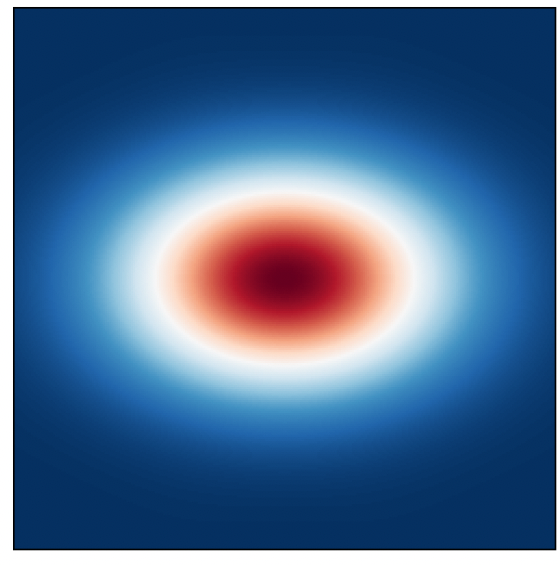}};
\node[inner sep=0pt] (bloch) at (8.75,1.)
    {\includegraphics[width=5.5cm]{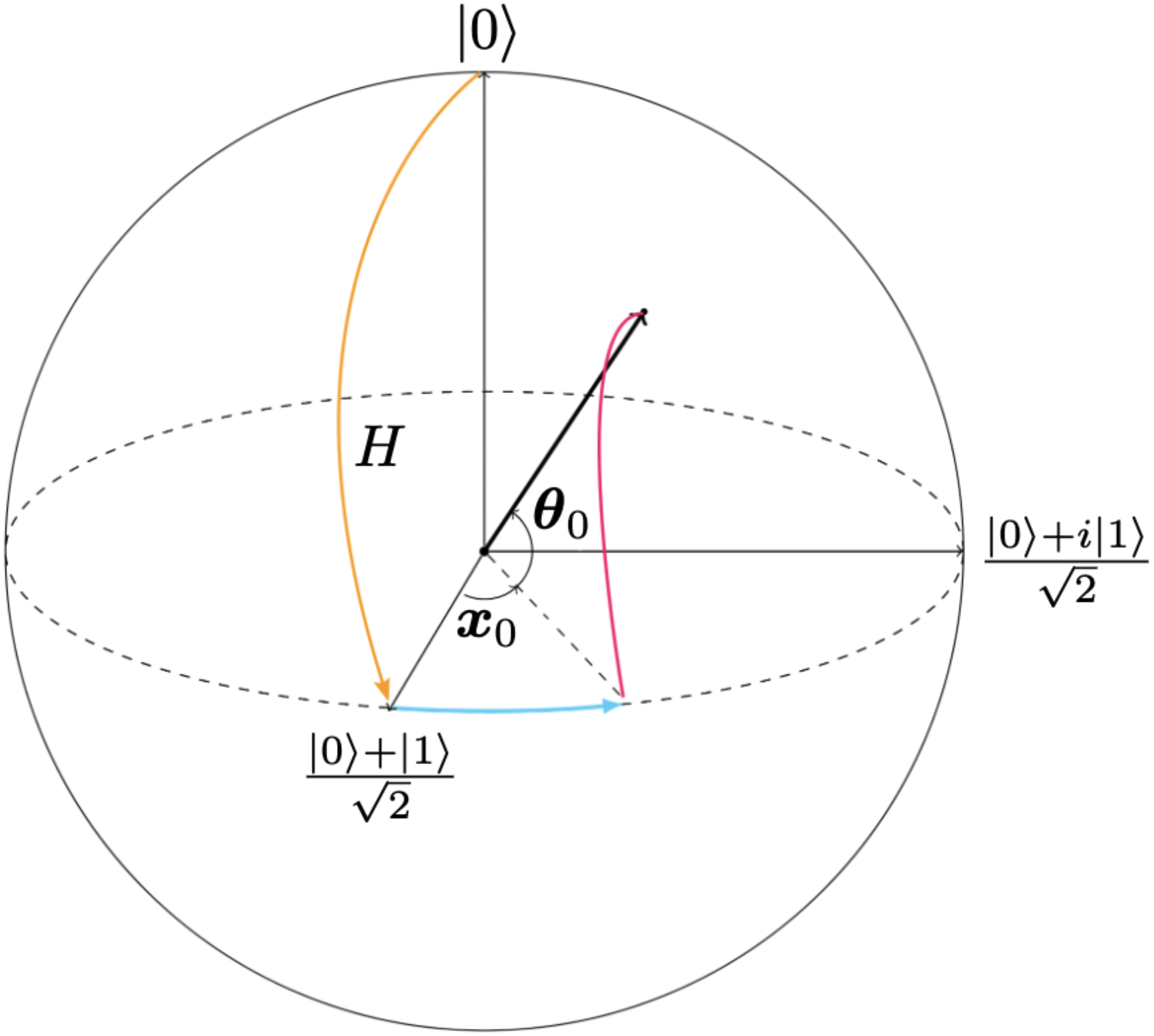}};
    \draw [decorate,thick, decoration = {brace}] (-5.85,2.6) --  (1.86,2.6);
    \draw [decorate,thick, decoration = {brace}] (-5.5,-2.7) --  (-4.5,-2.7);
    \draw [decorate,thick, decoration = {brace}] (-5.5,-2.25) --  (1.015,-2.25);
    \node at (-2,3.) { $d=1$ };
    \node at (-4.95,-2.4) { $l=1$ };
    \node at (-2.2,-1.95) { $l=2$ };
  \node at (-6.6,3.5) {\Large (a)};
  \node at (-6.6,-2) {\Large (c)};
  \node at (4.1,3.5) {\Large (b)};
  \node at (4.1,-2) {\Large (d)};
  \node at (4.05,-4.3) {\large $\boldsymbol{p}$};
  \node at (8.5,-4.3) {\large $\boldsymbol{p}$};
  \node at (6.1,-6.3) {\large $\boldsymbol{x}$};
  \node at (10.55,-6.3) {\large $\boldsymbol{x}$};
\end{tikzpicture}
  \caption{\textbf{Quantum Support Vector Machine Architectures.} 
   We implement quantum support vector machines (QSVMs) based on both qubit and continuous variable based quantum computing hardware.
   (a) The architecture of our qubit-based QSVMs. The embedding consists of repeated layers of Hadamard gates,
   data encoding $z$ rotations, parameterised $y$ rotations and entangling controlled $x$ rotations. 
   These layers are repeated until the entire event (a vector with 130 components) has been encoded. 
   We utilise 10 qubits QSVMs with 52 such layers (so the input data is encoded $10\times 52/130=4$ times). 
   (b) In the initial stages of the embedding, each qubit is mapped from the state $\ket{0}$ by a data dependent rotation,
   which can be represented on the Bloch sphere.
   (c) As described in Equation~\ref{eq:embed}, our embedding procedure in the continuous variable case
   consists of a variable number $l$ of alternating layers
   of displacement and nearest neighbour two mode squeezing operations (see Equations~\ref{eq:disp} and ~\ref{eq:squeeze} respectively).
   When $l=1$ only single mode displacement operations are employed, and the modes remain unentangled.
   The initial state $\ket{0}^{\otimes n}$ denotes the vacuum state of the system. 
   (d) 
   The effects of the squeezing and displacing operations on the vacuum state can be visualised by examining the resulting Wigner functions
   (the Wigner function of the vacuum is a Gaussian centred at the origin of phase space).
   Left: a displaced vacuum state; 
   Right: a squeezed vacuum state. The magnitude of the displacing and squeezing is determined by the input event, producing a data dependent encoded state. 
  }
\label{fig:1}
\end{center}
\end{figure*}
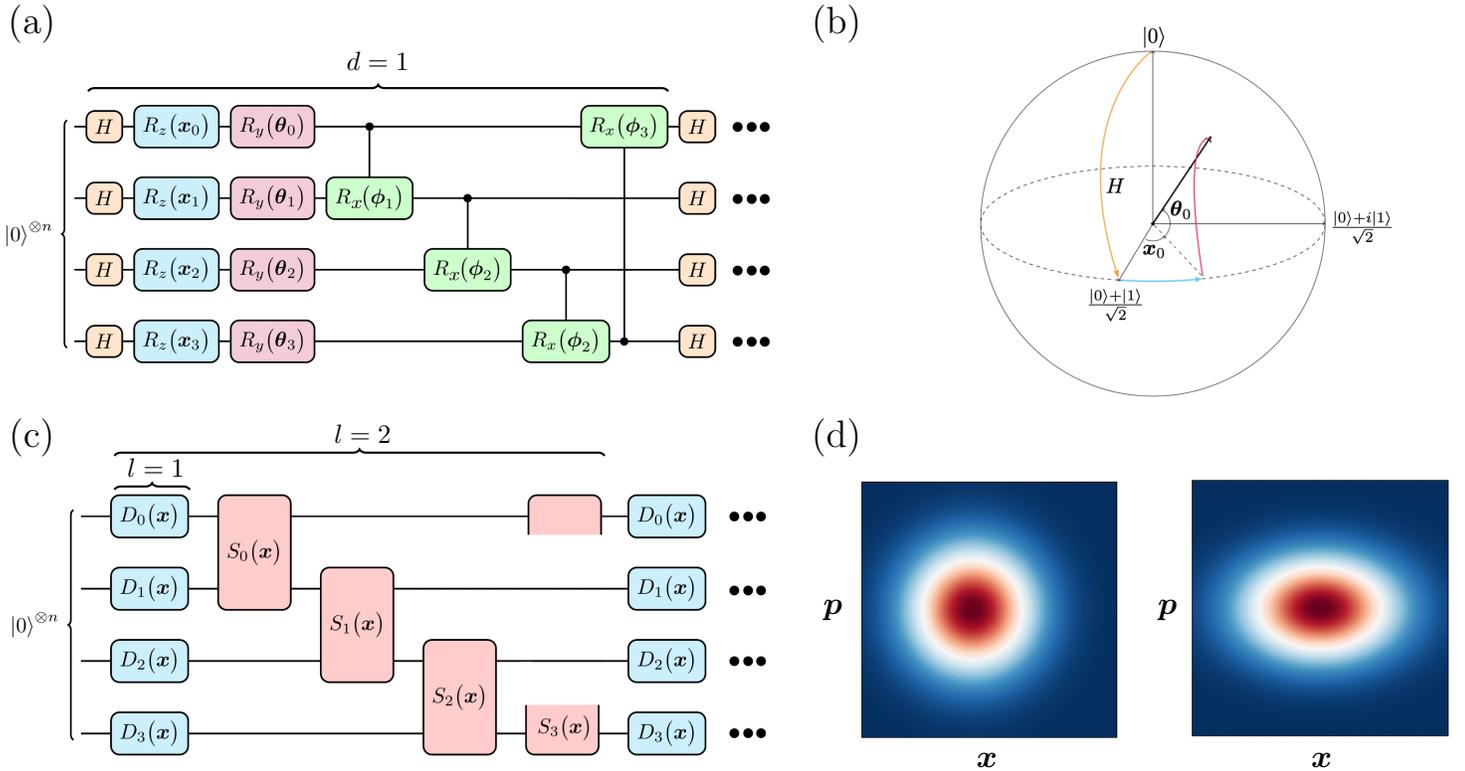

In this work we introduce ensembles of
boosted quantum support vector machines (QSVMs) as a technique for performing $B$ meson 
flavour tagging near the level of state-of-the-art classical algorithms. 
QSVMs are powerful classifiers which both have the potential to exponentially outperform their classical counterparts~\cite{liu2021rigorous}
and are capable of being implemented on near-term NISQ devices~\cite{heredge2021quantum},
making them attractive candidates to study in the search for practically useful applications of QML.
Indeed, QSVMs have already been employed in the analysis of
high energy physics data, for example to distinguish signal from background in 
neutral $B$ meson decays~\cite{heredge2021quantum} and $t\overline{t}H$ production~\cite{wu2021application}.
Here we apply them for the first time
to full-scale $B$ factory simulated data, with no simplifying assumptions about the number or type of decay 
products resulting from the collision, and including background derived from real experimental conditions,
in order to perform direct comparisons with the latest classical techniques.
This necessitates using QSVMs which accept 130 datapoints as input, the simulation of which involves extensive 
classical computing resources.
Performing such large-scale simulations is an important milestone on the journey to QML 
algorithms which genuinely outperform their classical counterparts in real world scenarios.

Our investigation of QSVMs for $B$ meson flavour tagging is performed by benchmarking two 
implementations: conventional qubit based QSVMs (see Figure~\ref{fig:1}(a,b)), and QSVMs based on the 
promising continuous variable (CV) quantum computing 
model~\cite{blance2021unsupervised,gottesman2001encoding,killoran2019continuous,gu2009quantum,bourassa2021blueprint} (Figure~\ref{fig:1}(c,d)).
In contrast to quantum computers built from qubits (as realised for example via superconducting circuits~\cite{jurcevic2021demonstration}, 
donor implantation in semiconductors~\cite{doi:10.1126/sciadv.1500707} and trapped ions~\cite{bruzewicz2019trapped})
CV quantum computers operate on bosonic modes, realised in quantum optical systems~\cite{gu2009quantum,bourassa2021blueprint}. 
Our decision to employ CV quantum computers is inspired by the observation that they can naturally emulate 
the highly successful radial basis 
function kernel~\cite{rbf} popularly employed in classical SVMs, which can be used as a starting point 
for developing more interesting quantum kernels.
Indeed, we find that QSVMs constructed from Gaussian operations in the CV picture  (CV-QSVMs)
can outperform QSVMs constructed in the usual qubit model out of single qubit rotations and 
entangling 2-qubit controlled rotations,
achieving tagging efficiencies of 29.2\% and 28.0\% respectively, approaching the 
peak tagging efficiency published using classical machine learning techniques (30.0\% using fast boosted decision
trees~\cite{flavourtagging}). 
Crucial to the success of the QSVMs are the twin applications of ensemble learning and boosting. 
By constructing ensembles of 200 QSVMs and averaging the results, while also boosting each individual QSVM
using the AdaBoost algorithm~\cite{adaboost}, we achieve drastic increases in performance over a single, non-boosted
QSVM (14.6\% and 14.0\% for qubit and continuous variable  QSVMs respectively).
Thus, boosted ensembles transform a weak quantum classifier into one which 
is commensurate with powerful classical ML techniques.
Moreover, our results are achieved while working solely with classically simulable QSVMs, 
severely restricting the available choices for the design of the CV-QSVMs.
By analysing the performance of the CV-QSVMs in a simplified setting resulting from a dimensionality reducing
PCA transformation~\cite{jolliffe2016principal} we argue that CV-QSVMs beyond the classically 
simulable regime may be able to outperform
the reported classical results, when quantum computing hardware becomes capable enough to run them.

%\large{{\textbf{3. Continuous Variable Quantum \\ \phantom{...} Support Vector Machines}}}
\section{Methods}
\subsection{Quantum Support Vector Machines}
Support vector machines (SVMs) are linear classifiers on data which has typically been non-linearly mapped into 
a high dimensional feature space~\cite{bottou2007support}.
Given two classes of data, an SVM attempts to find a hyperplane in the feature space which maximally separates them.
In a quantum support vector machine (QSVM) the mapping is into a quantum Hilbert space, $\x\mapsto\ket{\psi(\x)}\bra{\psi(\x)}$.  
QSVMs were one of the first QML algorithms to be introduced~\cite{havlivcek2019supervised}, 
and are thought to offer the potential for quantum advantage due to their ability to implement feature maps which make use of 
the exponentially large Hilbert spaces available to a quantum computer, hopefully making the embedding of the data into linearly separable 
subsets possible. 
Indeed it has been demonstrated that in principle QSVMs can offer an exponential speed-up over their classical counterparts
by efficiently detecting patterns equivalent to solving problems which are not thought to be classically solvable in polynomial time~\cite{liu2021rigorous},
although it remains unclear how often such dramatic benefits will be seen in practice on real-world data.
Despite the absence of robust theoretical guarantees, however, early work applying QSVMs to problems in high energy physics 
has shown that they can achieve results competitive with classical methods, at least on small-scale data~\cite{heredge2021quantum}.
Importantly, one never has to explicitly read out any (exponentially large) quantum states, as
an SVM does not explicitly utilise the embedding of individual datapoints into the embedding space, but rather 
only the pair-wise inner products between the embedded datapoints, information which is readily available to quantum computers.
Specifically, given an embedding $\psi$, a QSVM is a function only of
the \textit{kernel matrix}
\begin{equation}
  K\left(\x_i, \x_j\right) =  \abs{\bra{\psi\left(\x_i\right)}\ket{\psi\left(\x_j\right)}}^2  \label{eq:kernel}
\end{equation}

\noindent
over all pairs $\left(\x_i, \x_j\right)$ of training events.
Having used a quantum computer (or in our case, a simulation of a quantum computer) to calculate $K$, the 
rest of the procedure is an entirely classical algorithm implemented within many standard ML frameworks.
In this work we utilise the implementation of \texttt{scikit-learn}~\cite{scikit-learn}.\\

\begin{figure*}[ht]
 \begin{center}
 \includegraphics[width=0.95\textwidth]{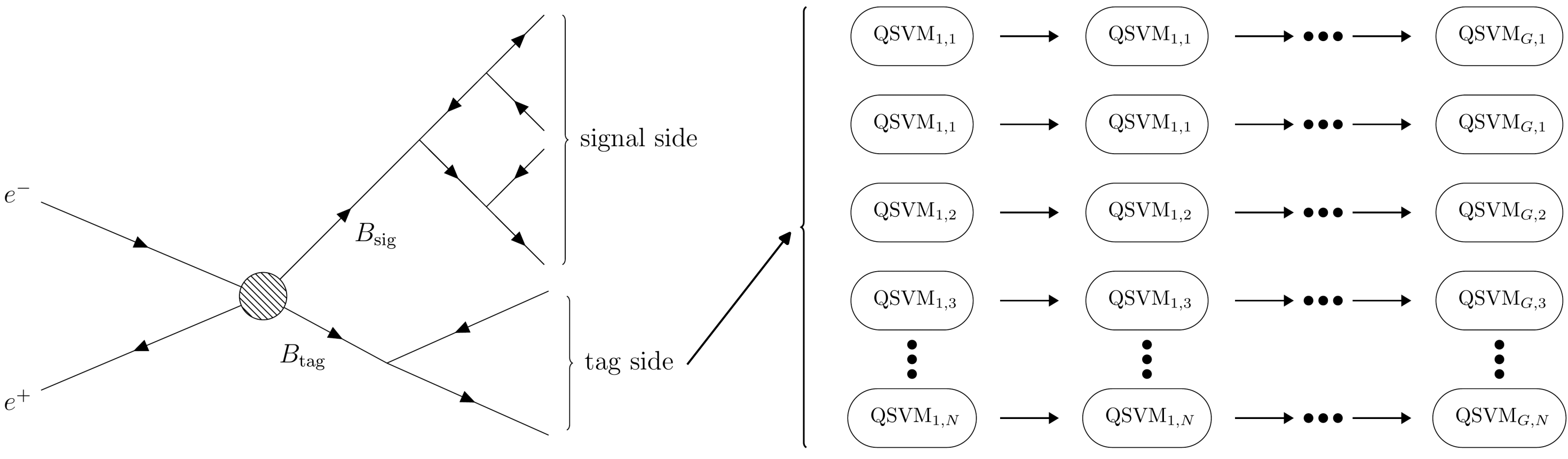}
  \caption{\textbf{Flavour tagging with boosted ensembles of quantum support vector machines.}
   The collision of $e^-e^+$ pairs at the $\Upsilon (4S)$ resonance can lead to the production of entangled ${B}^0\overline{B}{}^0$ pairs,
   one of which
   subsequently decays into a flavour agnostic $CP$ eigenstate (the \textit{signal side} meson) and the other to a 
   possibly flavour-specific state
   (the \textit{tag side} meson). The tag side decay products are inputted into a flavour tagging algorithm which attempts to determine
   the flavour of the parent (tag-side) $B$ meson. Due to the entangled state in which the pair of $B$ mesons were originally created 
   (see Equation~\ref{eq:entangled}), the flavour 
   of the signal side meson can then be inferred from the tag side flavour.
   For our flavour tagging algorithm we employ ensembles of $N$ quantum support vector machines (QSVMs) which are
   boosted with the AdaBoost algorithm~\cite{adaboost} for $G$ generations. At testing time, the flavour is predicted by a majority
   vote of the QSVMs, with the confidence $\abs{qr}$ determined by the margin of the vote. 
   We consider both QSVMs constructed from standard parameterised gates acting on qubits (see Figure~\ref{fig:1}(a)) 
   and QSVMs constructed from parametrised Gaussian operations acting on qumodes (Figure~\ref{fig:1}(c)).
   }
  \label{fig:ft_cartoon}
 \end{center}
\end{figure*}

The key component of a QSVM, then, is the map which embeds a datapoint $\x$
into the Hilbert space of the quantum computer.
Here we implement this map in two different ways, considering both circuits 
constructed from single qubit rotations and two qubit entangling gates in the standard qubit picture (see Figure~\ref{fig:1}(a,b)),
and circuits constructed from alternating layers of displacement and squeezing operations in the 
continuous variable (CV) picture of quantum computing~\cite{gottesman2001encoding} (see Figure~\ref{fig:1}(c,d)).
Explicitly, in the qubit case we have (for a circuit with $d$ layers)
\begin{align}
  \x \mapsto \ket{\psi(\x)}_{\mathrm{qubit}} = \bigotimes_{i=1}^d\bigg[ &\prod_{k=1}^n\left(CR_{x}^{k,k+1(\mathrm{mod}\ n)}(\phi_{ni+k})\right) \nonumber \\
  &\prod_{j=1}^n\left(R_y^j(\theta_{ni+j}) R_z^{j}(x_{ni+j})\right)H^{\otimes n}\bigg] \ket{0}^{\otimes n} \label{eq:qubit_embed}
\end{align}
where $H$ is the Hadamard gate, $R_z^{j}$ a $z$ rotation on the $j$th qubit (similarly for $R_y^{j}$) and
$CR_{x}^{a,b}$ a controlled $x$ rotation with control qubit $a$ and target qubit $b$.
We work wth $n=10$ qubits.
While our qubit based QSVMs are quite standard and similar to designs which have previously appeared in the literature~\cite{heredge2021quantum}, 
CV based QSVM architectures are comparatively understudied, and so we now describe those models
in detail.

%\subsection{Continuous Variable Quantum \phantom{.....} \phantom{} Support Vector Machines}
\subsection{Continuous Variable Quantum Support\\ Vector Machines}
Continuous variable quantum computers~\cite{gottesman2001encoding,killoran2019continuous,gu2009quantum,bourassa2021blueprint} have emerged as a candidate architecture for large-scale quantum computers developing in parallel 
to the mainstream qubit based systems,
and are defined by their use of \textit{continuous quantum variables}, quantum systems with continuous degrees of freedom.
Concretely, the fundamental units of a CV quantum computer are \textit{qumodes} (as opposed to qubits), which are states of the countably infinite 
dimensional bosonic Fock space $\mathcal{F}_{+}(\mathbb{C})$, with the total Hilbert space of an $n$ qumode CV quantum computer then
being $\mathcal{H}=\left(\mathcal{F}_{+}(\mathbb{C})\right)^{\otimes n}$.
Experimentally, such systems are most readily constructed within the framework of quantum optics~\cite{madsen2022quantum}, with each qumode corresponding 
to a different mode of quantised light, representable by a quantum harmonic oscillator.
The state of a single such qumode $\ket{\psi}_{\mathrm{qumode}}$ may be written in the photon number basis,
\begin{equation}
  \ket{\psi}_{\mathrm{qumode}} = \sum_{n=0}^\infty \alpha_n \ket{n}  \label{eq:qumodes}
\end{equation}
 
\noindent
with transitions between states of definite photon number implemented by the usual harmonic oscillator annihilation and 
creation operators
${a}$ and ${a}^\dagger$:
\begin{equation}
  {a} \ket{n} = \sqrt{n}\ket{n-1};\qquad {a}^\dagger \ket{n} = \sqrt{n+1}\ket{n+1}   \label{eq:create}
\end{equation}

\noindent
Our continuous variable quantum support vector machines (CV-QSVMs) involve embedding the event data $\x$ into states in $\mathcal{H}$
via parametrised operators built from $a$ and $a^\dagger$.
In particular, we employ multi-mode displacement and squeezing operations, respectively defined by

\begin{equation}
  D(\boldsymbol{x}) = \prod_{i=0}^{n}D_i(\x) = \prod_{i=0}^{n}e^{\beta\left(x_i ^*  a_i^\dagger\ -\ x_i a_i\right)} \\  \label{eq:disp}
\end{equation}
\begin{equation}
  S(\boldsymbol{x}) = \prod_{i=0}^{n}S_i(\x)= \prod_{i=0}^{n}e^{\gamma\left(x_i ^* a_ia_{i+1}\ +\ x_i a_i^\dagger a_{i+1}^\dagger\right)}  \label{eq:squeeze}
\end{equation}

\noindent
where $a_i\ (a_i^\dagger)$ is the annihilation (creation) operator acting on the $i$th qumode, and $\beta$ and $\gamma$ are hyperparameters which we set to be $\beta=\gamma=0.1$.
%The states $D(\x)\ket{0}^{\otimes n}$ and $S(\x)\ket{0}^{\otimes n}$ are referred to as \textit{displaced states} and \textit{squeezed states} respectively.
The action of the squeezing and displacement operators may be visualised 
by considering their effect on the Wigner functions $W(x,p)$ of the states $\ket{\psi}$, where we recall
\begin{equation}
  W(x,p) = \frac{2}{\pi} \int dy \ e^{4iyp} \bra{x-y}\ket{\psi}\bra{\psi}\ket{x+y} \label{eq:wigner}
\end{equation}
with $\ket{x}$ and $\ket{p}$ being eigenstates of the position and momentum operator, respectively
$\hat{x} = \left(\hat{a}^\dagger + \hat{a}\right)/{\sqrt{2}}$ and $\hat{p} = i\left(\hat{a}^\dagger - \hat{a}\right)/{\sqrt{2}}$. 
A quasi-probability distribution on phase space, the Wigner functions of a single mode (initially in the vacuum state) subjected to displacement and 
squeezing are shown in Figure~\ref{fig:1}(d), explaining the nomenclature ``displacement'' and ``squeezing''.\\

As previously discussed, the contribution of the quantum computer in the QSVM algorithm is to carry out the mapping $\x \mapsto \ket{\psi(\x)}$ which embeds a datapoint $\x$
into the Hilbert space of the quantum computer, and to take inner products of embedded datapoints.
In the CV setting we implement this mapping with circuits constructed from alternating layers of displacement and squeezing operations,

\begin{equation}
  \x \mapsto \ket{\psi(\x)}_{\mathrm{qumode}} = \underbrace{S(\boldsymbol{x})D(\boldsymbol{x})\cdots S(\boldsymbol{x})D(\boldsymbol{x})}_{l\ \text{operations}}\ket{0}^{\otimes n} \label{eq:embed}
\end{equation}

\noindent
for various values of $l$ (see Figure~\ref{fig:1}(c)). 
We simulate the circuits within the framework of \texttt{bosonic-qiskit}~\cite{stavenger2022bosonic}, which allows us to represent a qumode 
as a set of qubits. Although each qumode is formally a state of an infinite dimensional Hilbert space (Equation~\ref{eq:qumodes}), 
we can in practice truncate these spaces at some dimension $D$ and represent each qumode with $\log_2 D$  qubits.
Further details may be found in Ref.~\cite{stavenger2022bosonic}.
In this work we take $D=8$, thus assigning three physical qubits to each logical qumode.

\subsection{$B$ Meson Flavour Tagging}
$B^0-\overline{B}{}^0$ pairs are routinely created by asymmetric electron-positron collisions 
at the $\Upsilon (4S)$ resonance at 
the Belle-II experiment (see Figure~\ref{fig:ft_cartoon})~\cite{bfactories}
in order to investigate some of the less well understood aspects of the Standard Model, including
$CP$ violating effects~\cite{gronau2005precise,beneke2005corrections,abe2001observation} 
and the possibility of of lepton flavour non-universality~\cite{capdevila2018patterns,aebischer2020b,aaij2019search}.
Flavour tagging is the process of determining the quark flavour content of each of the $B$ mesons, and is a critical step of
the analysis of many experiments which probe $CP$ violation and $ B^0-\overline{B}{}^0 $ mixing~\cite{flavourtagging}.
It utilises the 
maximally entangled form of the coherent state $\ket{\Psi}$ prepared by the $e^+e^- \to \Upsilon (4S) \to B^{0}\overline{B}{}^0 $
transition, 
\begin{equation}
  \ket{\Psi} = \frac{1}{\sqrt{2}} \left(\ket{B^{0}\overline{B}{}^0} - \ket{\overline{B}{}^{0}B^0} \right) \label{eq:entangled}
\end{equation} 
and the resulting perfect anticorrelation of the flavours of the two mesons, 
to determine the flavour of a  $B$ meson which is involved in otherwise ambiguous (neutral) decays.
For example, to investigate the asymmetry 
\begin{equation}
  A_{CP}=\frac{\Gamma(\overline{B}{}^0\to \pi^0\pi^0) - \Gamma({B}^0\to \pi^0\pi^0)}{\Gamma(\overline{B}{}^0\to \pi^0\pi^0) + \Gamma({B}^0\to \pi^0\pi^0)} \label{eq:decay}
\end{equation} 
in the decay rate $\Gamma$ of $ B^0$s and $ \overline{B}{}^0 $s to the $CP$ eigenstate $\pi^0\pi^0$,
one can infer the flavour of the $B$ meson which underwent the decay to the neutral final state $\pi^0\pi^0$ (the \textit{signal side} meson)
by determining the flavour of the other $B$ meson (the \textit{tag side} meson)
and then invoking the anti-correlation of the flavour of the $B$ mesons at the point of the creation.
(see Figure~\ref{fig:1} and Ref.~\cite{bfactories}).
The accuracy with which one can carry out the flavour tagging process directly impacts upon the precision of
measurements of $CP$ asymmetries such as in Equation~\ref{eq:decay}~\cite{bfactories}.
Improved flavour tagging algorithms therefore allow for increasingly sharp tests of SM predictions, and may 
lead to the discovery of New Physics beyond the SM~\cite{gronau2005precise,beneke2005corrections}.

\begin{figure}
\begin{center}
	\begin{tikzpicture}
	\begin{feynman}
	\vertex (a1) {\(\overline b\)};

	\vertex[right=3cm of a1] (a5);
	\vertex[right=2cm of a5] (a6) {\(\)};
	\vertex[below=2em of a6] (a7) {\(\overline{c}\)};
	
	\vertex[below=2em of a1] (b1) {\(d\)};
	\vertex[below=2em of a5] (b5);
	\vertex[below=4em of a6] (b6) {\(d\)};
	
	\vertex[above=2em of a6] (c1) {\(e^+\)};
	\vertex[above=2em of c1] (c3) {\(\nu_e\)};
	\vertex at ($(c1)!0.5!(c3) - (1.5cm, 0)$) (c2);
	
	\diagram* {
		{[edges=fermion]
			(b1) --  (b5) -- (b6),
			(a7) --  (a5) -- (a1),
		},

		(c1) -- [fermion] (c2) -- [fermion] (c3),
		(a5) -- [boson, edge label=\(W^{+}\)] (c2),
	};
	
	\draw [decoration={brace}, decorate] (b1.south west) -- (a1.north west)
	node [pos=0.5, left] {\(B^{0}\)};
	%\draw [decoration={brace}, decorate] (a6.north east) -- (b5.south east)
	%node [pos=0.5, right] {\(\pi^{+}\)};
	%\end{feynman}
	%\end{tikzpicture}
	%\begin{tikzpicture}
	%\begin{feynman}
 \vertex[below=3.5cm of a1] (ba1) {\( b\)};

	\vertex[right=3cm of ba1] (ba5);
	\vertex[right=2cm of ba5] (ba6) {\(\)};
	\vertex[below=2em of ba6] (ba7) {\({c}\)};
	
  \vertex[below=2em of ba1] (bb1) {\(\overline{d}\)};
	\vertex[below=2em of ba5] (bb5);
  \vertex[below=4em of ba6] (bb6) {\(\overline{d}\)};
	
	\vertex[above=2em of ba6] (bc1) {\(e^-\)};
  \vertex[above=2em of bc1] (bc3) {\(\overline{\nu}_e\)};
	\vertex at ($(bc1)!0.5!(bc3) - (1.5cm, 0)$) (bc2);
	
	\diagram* {
		{[edges=fermion]
			(bb6) --  (bb5) -- (bb1),
			(ba1) --  (ba5) -- (ba7),
		},

		(bc3) -- [fermion] (bc2) -- [fermion] (bc1),
		(ba5) -- [boson, edge label=\(W^{-}\)] (bc2),
	};
	
	\draw [decoration={brace}, decorate] (bb1.south west) -- (ba1.north west)
    node [pos=0.5, left] {\(\overline{B}{}^{0}\)};
	%\draw [decoration={brace}, decorate] (a6.north east) -- (b5.south east)
	%node [pos=0.5, right] {\(\pi^{+}\)};
	\end{feynman}
	\end{tikzpicture}
   \captionof{figure}{\label{fig:ft}\textbf{Primary leptons.}
   Top: in the semileptonic decay $B^0 \mapsto Xl^{+}\nu_l$ (for some hadron $X$, here a $D^-$ meson) the positive charge of the so-called primary lepton
   $l$ unambiguously identifies the flavour of the original $B$ meson. Bottom: similarly, in the charge conjugate process
   $\overline{B}{}^0 \mapsto Xl^{-}\overline{\nu}_l$ the charge of the primary lepton again determines the flavour of the parent $B$ meson. 
   }
\end{center}
\end{figure}
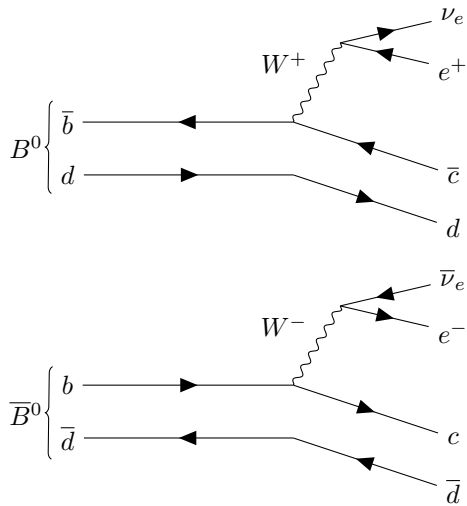

As an example of flavour tagging we consider the semileptonic decay $B^0 \mapsto Xl^{+}\nu_l$ for some hadron $X$
(see Figure~\ref{fig:ft}). In this case the positive charge of the \textit{primary lepton} $l^{+}$ unambiguously determines the flavour of 
the parent $B$ meson (c.f. the charge conjugate process $\overline{B}{}^0 \mapsto {X}l^{-}\overline{\nu}{}_l$, also shown in 
Figure~\ref{fig:ft}).
Unfortunately the charge of a resulting lepton alone is in general insufficient to determine the flavour,
as for example the hadron $X$ may itself decay, emiting a \textit{secondary lepton} of the opposite charge to the primary lepton.
Such a secondary decay product will however have a different momentum distribution to that of the primary lepton,
and by combining the momenta and particle type of all of the decay products we can hope to infer the flavour with high probability.
In practice this has been most successfully accomplished by feeding all of the available information from the event into a 
classical ML model, with both fast boosted decision trees and deep neural networks having been employed successfully~\cite{flavourtagging,ftdnn,bfactories}.

Given an event $\mathcal{E}$, a flavour tagger customarily outputs a prediction $qr\in [-1,1]$, with $q\in \{-1,1\}$ 
denoting the predicted flavour, and $r=\abs{qr}$ the confidence of the prediction.
The performance of a flavour tagger is typically measured by its effective tagging efficiency $\epsilon_{\mathrm{eff}}$ on a set of test events,
defined as
\begin{equation}
  \epsilon_{\mathrm{eff}} = \sum_i \epsilon_i \left(1-2w_i\right)^2 \label{eq:tag}
\end{equation}
\noindent
where $i$ indexes mutually orthogonal bins of events corresponding to predictions of various levels of confidence,
and $w_i$ is the fraction of events in the $i$th bin which are misclassified by the flavour tagger.
In this work we employ seven bins, consistent with the Belle-II convention~\cite{bfactories,flavourtagging}.
The definition of tagging efficiency given in Equation~\ref{eq:tag} is employed as the figure of merit of flavour tagging algorithms 
(rather than, say, the raw accuracy) due to the observation~\cite{bfactories} 
that the statistical uncertainty $\sigma$ of $CP$ asymmetric measurements scales approximately as $\sigma\propto \epsilon_{\mathrm{eff}}^{-1/2}$.
As a major goal of flavour tagging is to serve as a step in the analysis of $CP$ violating decays,
maximising the tagging efficiency becomes the primary goal when training a classifier. 
Previously, (classical) fast boosted decision trees and deep neural networks have been employed to achieve 
effective tagging efficiencies of $30.0\%$  and $28.8\%$ respectively~\cite{flavourtagging}.\\

\section{Results and Discussion}
\begin{figure}[t]
 %\centering
\hspace{-5mm}
 \begin{tikzpicture}[line cap=round]
  \begin{axis}[
      domain=-1.5:3,samples=80,
      xmin=0, xmax=106.33,
      ymin=0, ymax=1.02,
      anchor=origin,x=0.0632cm,y=5.8cm, % coincide with TikZ coordinates
      ytick style={draw=none},
      xtick style={draw=none},
      title={Qubit Based, Single QSVM},
      %axis equal image, % fit TikZ image
      %grid=both,
      xlabel=Training Events $(\times 1000)$,
      ylabel=Tagging Efficiency,
      %ylabel=$y$,
    legend style={at={(0.495,0.22)},anchor=west}
    ]
    %\addplot[thick,blue, scatter,scatter/classes={a={blue},b={blue},c={blue},d={blue},e={blue}}] table {testdata.txt};
    %\addplot[thick,blue, scatter,scatter/classes={a={blue}}] table {testdata.txt};
    \addplot[thick,Cerulean,skip coords between index={10}{20}, scatter,scatter/classes={a={Cerulean}}] table {overfitdata.txt};
    \addplot[thick,WildStrawberry,skip coords between index={0}{10}, scatter,scatter/classes={b={WildStrawberry}}] table {overfitdata.txt};
    \legend{Training Data, Test Data}
  \end{axis}
\end{tikzpicture}
  \caption{\label{fig:overfit}\textbf{Overfitting in QSVMs.} The tagging efficiencies on training and test data for  a 10 qubit, 13 layer QSVM
  as in Figure~\ref{fig:1}(a) as a function of the training set size. We observe massive overfitting, with the QSVM able to almost perfectly 
  classify a training set of 100,000 events, while barely generalising at all to the test data. Modifying the regularisation hyperparameter 
  of the QSVM has only a minor effect on the test performance (see  Figure~\ref{fig:reg}, Appendix A). Due to the expensive scaling of SVMs with the size of the training set,
   it is infeasible to combat this by increasing the size of the training set far beyond 100,000 }
\end{figure}
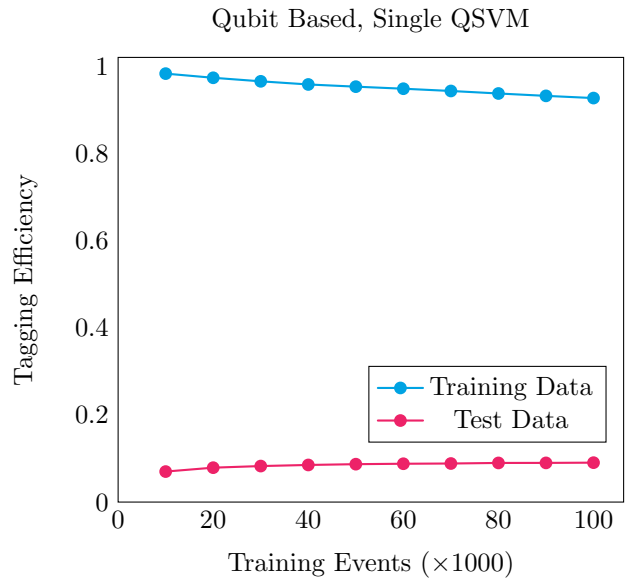

The tagging efficiencies obtained on the training and test data for a generic 10 qubit 52  layer QSVM as depicted in Figure~\ref{fig:1}(a)
are shown in Figure~\ref{fig:overfit} as a function of the number of events used in the training process. 
We observe severe overfitting, with the tagging efficiency on the unseen test data failing to rise above 12\%
(c.f. the classical state of the art value of $\sim 30\%$), depite extremely high performance on the training data. 
Remarkably, the high expressibility of QSVMs~\cite{srikumar2022kernel} allows it to 
exploit the $2^{n_{\mathrm{qubits}}}=1024$ dimensional Hilbert space to 
almost perfectly fit 
entire training sets of 100,000 examples without successfully generalising.
We find that varying the strength of the QSVM regularisation is of limited help in this regard (see Figure~\ref{fig:reg}, Appendix A).
This failure to generalise beyond the training set can also be seen as a reflection of the difficulty of 
$B$ meson flavour tagging as a classification task, 
with the high-performance classifiers developed to date employing large training datasets 
of $\sim 10^7$ events~\cite{flavourtagging,ftdnn,bfactories}.
Unfortunately, SVMs scale poorly with the size  of the training dataset (at least quadratically~\cite{bottou2007support}) 
which makes training an SVM, either classical or quantum, on a comparable number of events infeasible. \\

\pgfplotsset{scaled y ticks=false}

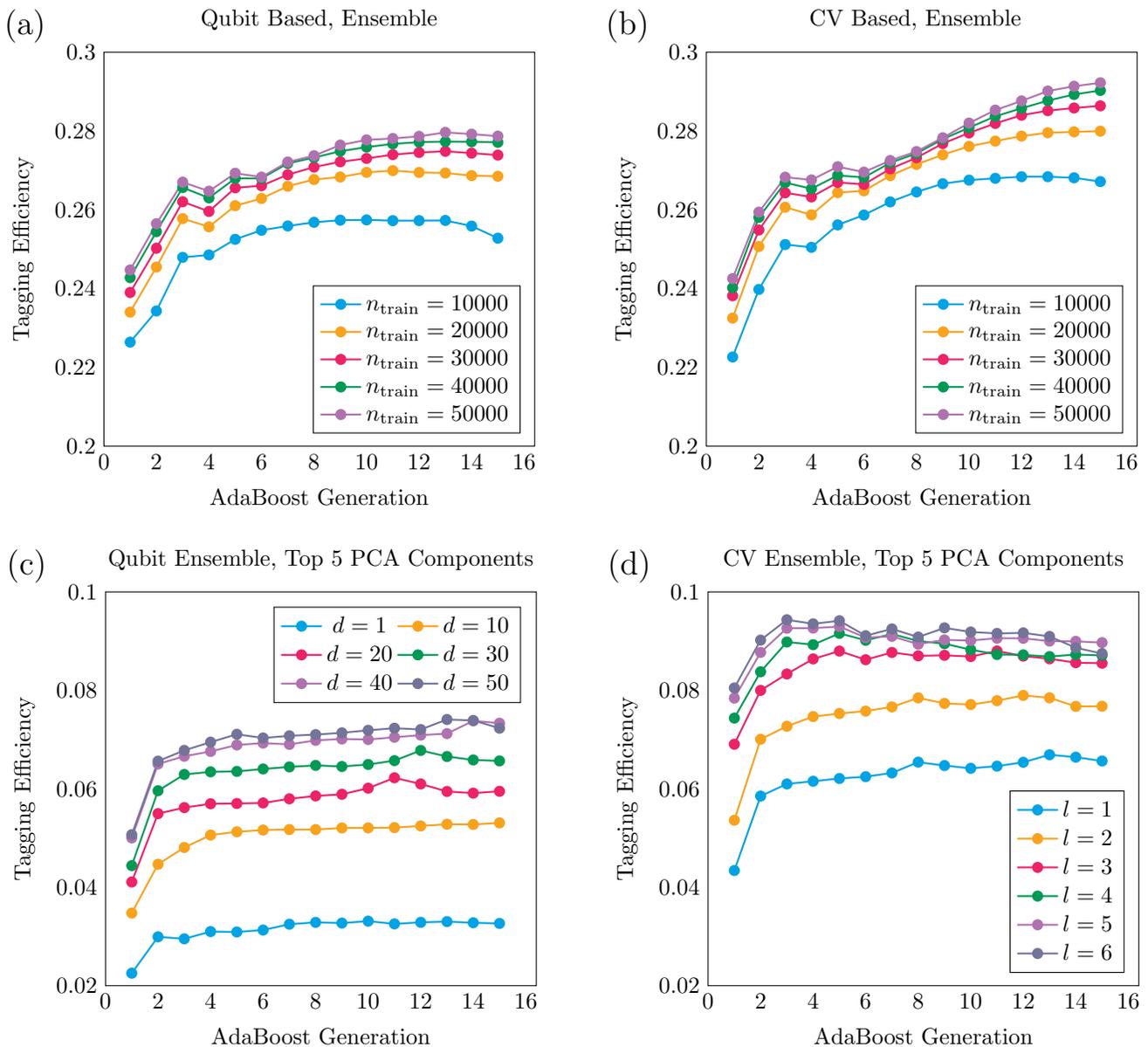
\begin{figure*}[ht]
 \begin{tikzpicture}[line cap=round]
  \node at (-1.2,6.4) {\Large (a)};
  \begin{axis}[
      domain=-1.5:3,samples=80,
      xmin=0, xmax=16.4,
      ymin=0.2, ymax=0.3,
      anchor=origin,x=0.4cm,y=59.994cm, % coincide with TikZ coordinates
      ytick style={draw=none},
      xtick style={draw=none},
      %axis equal image, % fit TikZ image
      %grid=both,
      title={Qubit Based, Ensemble},
      xlabel=AdaBoost Generation,
      ylabel=Tagging Efficiency,
      yticklabel style={
     /pgf/number format/precision=3,
     /pgf/number format/fixed},
     legend columns=1,
      %ylabel=$y$,
    legend style={at={(0.485,0.22)},anchor=west}
    ]
    %\addplot[thick,blue, scatter,scatter/classes={a={blue},b={blue},c={blue},d={blue},e={blue}}] table {testdata.txt};
    %\addplot[thick,blue, scatter,scatter/classes={a={blue}}] table {testdata.txt};
    \addplot[thick,Cerulean,skip coords between index={15}{105}, scatter,scatter/classes={a={Cerulean}}] table {circuitdata.txt};
    \addplot[thick,YellowOrange,skip coords between index={0}{15},skip coords between index={30}{105}, scatter,scatter/classes={b={YellowOrange}}] table {circuitdata.txt};
    \addplot[thick,WildStrawberry,skip coords between index={0}{30},skip coords between index={45}{105}, scatter,scatter/classes={c={WildStrawberry}}] table {circuitdata.txt};
    \addplot[thick,ForestGreen,skip coords between index={0}{45},skip coords between index={60}{105}, scatter,scatter/classes={d={ForestGreen}}] table {circuitdata.txt};
    \addplot[thick,Orchid,skip coords between index={0}{60},skip coords between index={75}{105}, scatter,scatter/classes={e={Orchid}}] table {circuitdata.txt};
    \addplot[thick,CadetBlue,skip coords between index={0}{75},skip coords between index={90}{105}, scatter,scatter/classes={f={CadetBlue}}] table {circuitdata.txt};
    %\addplot[thick,Gray,skip coords between index={0}{90}, scatter,scatter/classes={g={Gray}}] table {pca_data.txt};
    \legend{$n_\mathrm{train}=10000$, $n_\mathrm{train}=20000$, $n_\mathrm{train}=30000$, $n_\mathrm{train}=40000$, $n_\mathrm{train}=50000$ }
    %\legend{$l=1$,$l=2$,$l=3$,$l=4$,$l=5$,$l=6$,RBF}
  \end{axis}
     %\node at (3.3,0.2) {\scriptsize 200 CV-QSVM Ensemble, 50000 Training Events};
     %\node at (3.865,5.5) {Top 5 PCA Components};
     %\node at (5.,0.5) {\scriptsize 50000 Training Events};
     %\node at (4.825,0.2) {\scriptsize 200 CV-QSVM Ensemble};
\end{tikzpicture}
    \hspace{6mm}
 \begin{tikzpicture}[line cap=round]
  \node at (-1.2,6.4) {\Large (b)};
  \begin{axis}[
      domain=-1.5:3,samples=80,
      xmin=0, xmax=16.4,
      ymin=0.2, ymax=0.3,
      anchor=origin,x=0.4cm,y=60cm, % coincide with TikZ coordinates
      ytick style={draw=none},
      xtick style={draw=none},
      title={CV Based, Ensemble\phantom{p}},
      %axis equal image, % fit TikZ image
      %grid=both,
      xlabel=AdaBoost Generation,
      ylabel=Tagging Efficiency,
      %ylabel=$y$,
    legend style={at={(0.485,0.22)},anchor=west}
    ]
    %\addplot[thick,blue, scatter,scatter/classes={a={blue},b={blue},c={blue},d={blue},e={blue}}] table {testdata.txt};
    %\addplot[thick,blue, scatter,scatter/classes={a={blue}}] table {testdata.txt};
    \addplot[thick,Cerulean,skip coords between index={15}{75}, scatter,scatter/classes={a={Cerulean}}] table {data.txt};
    \addplot[thick,YellowOrange,skip coords between index={0}{15},skip coords between index={30}{75}, scatter,scatter/classes={b={YellowOrange}}] table {data.txt};
    \addplot[thick,WildStrawberry,skip coords between index={0}{30},skip coords between index={45}{75}, scatter,scatter/classes={c={WildStrawberry}}] table {data.txt};
    \addplot[thick,ForestGreen,skip coords between index={0}{45},skip coords between index={60}{75}, scatter,scatter/classes={d={ForestGreen}}] table {data.txt};
    \addplot[thick,Orchid,skip coords between index={0}{60}, scatter,scatter/classes={e={Orchid}}] table {data.txt};
    \legend{$n_\mathrm{train}=10000$, $n_\mathrm{train}=20000$, $n_\mathrm{train}=30000$, $n_\mathrm{train}=40000$, $n_\mathrm{train}=50000$ }
  \end{axis}
\end{tikzpicture}
\phantom{.}\\
\phantom{.}\\
 \begin{tikzpicture}[line cap=round]
   \hspace{-1.5mm}
  \node at (-1.2,6.4) {\Large (c)};
  \begin{axis}[
      domain=-1.5:3,samples=80,
      xmin=0, xmax=16.4,
      ymin=0.02, ymax=0.1,
      anchor=origin,x=0.4cm,y=75cm, % coincide with TikZ coordinates
      ytick style={draw=none},
      xtick style={draw=none},
      %axis equal image, % fit TikZ image
      %grid=both,
      title={Qubit Ensemble, Top 5 PCA Components},
      xlabel=AdaBoost Generation,
      ylabel=Tagging Efficiency,
      ytick={0.02, 0.04, 0.06, 0.08, 0.10},
      yticklabel style={
      /pgf/number format/precision=2,
      /pgf/number format/fixed},
     legend columns=2,
      %ylabel=$y$,
    legend style={at={(0.39,0.84)},anchor=west}
    ]
    %\addplot[thick,blue, scatter,scatter/classes={a={blue},b={blue},c={blue},d={blue},e={blue}}] table {testdata.txt};
    %\addplot[thick,blue, scatter,scatter/classes={a={blue}}] table {testdata.txt};
    \addplot[thick,Cerulean,skip coords between index={15}{105}, scatter,scatter/classes={a={Cerulean}}] table {pca_qubits.txt};
    \addplot[thick,YellowOrange,skip coords between index={0}{15},skip coords between index={30}{105}, scatter,scatter/classes={b={YellowOrange}}] table {pca_qubits.txt};
    \addplot[thick,WildStrawberry,skip coords between index={0}{30},skip coords between index={45}{105}, scatter,scatter/classes={c={WildStrawberry}}] table {pca_qubits.txt};
    \addplot[thick,ForestGreen,skip coords between index={0}{45},skip coords between index={60}{105}, scatter,scatter/classes={d={ForestGreen}}] table {pca_qubits.txt};
    \addplot[thick,Orchid,skip coords between index={0}{60},skip coords between index={75}{105}, scatter,scatter/classes={e={Orchid}}] table {pca_qubits.txt};
    \addplot[thick,CadetBlue,skip coords between index={0}{75},skip coords between index={90}{105}, scatter,scatter/classes={f={CadetBlue}}] table {pca_qubits.txt};
    %\addplot[thick,Gray,skip coords between index={0}{90}, scatter,scatter/classes={g={Gray}}] table {pca_data.txt};
    \legend{$d=1$,$d=10$,$d=20$,$d=30$,$d=40$,$d=50$}
    %\legend{$l=1$,$l=2$,$l=3$,$l=4$,$l=5$,$l=6$,RBF}
  \end{axis}
\end{tikzpicture}
\quad
\hspace{1mm}
 \begin{tikzpicture}[line cap=round]
  \node at (-1.2,6.4) {\Large (d)};
  \begin{axis}[
      domain=-1.5:3,samples=80,
      xmin=0, xmax=16.4,
      ymin=0.02, ymax=0.1,
      anchor=origin,x=0.4cm,y=75cm, % coincide with TikZ coordinates
      ytick style={draw=none},
      xtick style={draw=none},
      %axis equal image, % fit TikZ image
      %grid=both,
      title={CV Ensemble, Top 5 PCA Components},
      xlabel=AdaBoost Generation,
      ylabel=Tagging Efficiency,
      ytick={0.02, 0.04, 0.06, 0.08, 0.1},
      yticklabel style={
     /pgf/number format/precision=3,
     /pgf/number format/fixed},
     legend columns=1,
      %ylabel=$y$,
    legend style={at={(0.70,0.264)},anchor=west}
    ]
    %\addplot[thick,blue, scatter,scatter/classes={a={blue},b={blue},c={blue},d={blue},e={blue}}] table {testdata.txt};
    %\addplot[thick,blue, scatter,scatter/classes={a={blue}}] table {testdata.txt};
    \addplot[thick,Cerulean,skip coords between index={15}{105}, scatter,scatter/classes={a={Cerulean}}] table {pca_data.txt};
    \addplot[thick,YellowOrange,skip coords between index={0}{15},skip coords between index={30}{105}, scatter,scatter/classes={b={YellowOrange}}] table {pca_data.txt};
    \addplot[thick,WildStrawberry,skip coords between index={0}{30},skip coords between index={45}{105}, scatter,scatter/classes={c={WildStrawberry}}] table {pca_data.txt};
    \addplot[thick,ForestGreen,skip coords between index={0}{45},skip coords between index={60}{105}, scatter,scatter/classes={d={ForestGreen}}] table {pca_data.txt};
    \addplot[thick,Orchid,skip coords between index={0}{60},skip coords between index={75}{105}, scatter,scatter/classes={e={Orchid}}] table {pca_data.txt};
    \addplot[thick,CadetBlue,skip coords between index={0}{75},skip coords between index={90}{105}, scatter,scatter/classes={f={CadetBlue}}] table {pca_data.txt};
    %\addplot[thick,Gray,skip coords between index={0}{90}, scatter,scatter/classes={g={Gray}}] table {pca_data.txt};
    \legend{$l=1$,$l=2$,$l=3$,$l=4$,$l=5$,$l=6$}
    %\legend{$l=1$,$l=2$,$l=3$,$l=4$,$l=5$,$l=6$,RBF}
  \end{axis}
     %\node at (3.3,0.2) {\scriptsize 200 CV-QSVM Ensemble, 50000 Training Events};
     %\node at (3.865,5.5) {Top 5 PCA Components};
     %\node at (5.,0.5) {\scriptsize 50000 Training Events};
     %\node at (4.825,0.2) {\scriptsize 200 CV-QSVM Ensemble};
\end{tikzpicture}
  \caption{\label{fig:results}\textbf{Flavour tagging efficiencies.}
   (a) The tagging efficiencies for ensembles of 200 10-qubit QSVMs with depth $d=52$, as depicted in 
   Figure~\ref{fig:1}(a), indexed by the number $n_{\mathrm{train}}$ of training event seen by each QSVM 
   and as a function of the number of boosting iterations. 
   By switching to this ensemble technique we observe a massive increase in performance relative to the case 
   of a single QSVM (cf Figure~\ref{fig:overfit} and Figure~\ref{fig:esize} in Appendix A).
   Moreover, boosting with AdaBoost~\cite{adaboost} provides a further significant improvement over 
   the vanilla QSVM ensemble.
   (b) Similarly, we evaluate 
   the performance of an ensemble of 200 boosted CV-QSVMs (with $l=1$, see Equation~\ref{eq:embed}) 
   throughout the boosting process. 
   With a peak tagging efficiency of 29.2\%, the CV-QSVMs are competitive with state-of-the-art 
   classical ML techniques~\cite{flavourtagging, bfactories}.
   Classically simulating an ensemble of CV-QSVMs with $l>1$ on the full 130 dimensional data is 
   computationally intractable.
   Additional plots showing the effect of changing the number of CV-QSVMs in the ensemble
   and the binning strategy used for the flavour tagging (see Equation~\ref{eq:tag})
   may be found in Figures~\ref{fig:esize} and~\ref{fig:bins} in Appendix A.
   (c, d) In order to investigate the performance of deeper QSVMs we consider 
   a reduced dataset consisting of the top 5 PCA components
   of the 130 dimensional input data. 
   In both the qubit and CV cases we find increasing the depth of the circuits highly beneficial.
Similar plots calculated using QSVMs with access to various numbers of PCA components are 
shown in Figure~\ref{fig:pca_comps}.
   %If similar benefits are found for the full scale CV-QSVMs of part (b), CV-QSVMs may surpass the 
   %reported performance
   %of the currently employed classical techniques.
   In all cases we test on 50,000 events.
   }
\end{figure*}

In order to mitigate this issue
we train large ensembles of $N$ QSVMs, which are individually trained on a
manageable number $(\sim 10^4)$ of events (see Figure \ref{fig:ft_cartoon}). 
At testing time each QSVM outputs its predicted $qr$ value for the test events,
and the final output of the global classifier is taken to be the average value over the ensemble.
By taking $N=200$ we ensure that the total number of events seen by the ensemble classifier is comparable to that of the classical algorithms
(see Figure~\ref{fig:esize} in Appendix A for the effect of varying $N$).
Additionally, we employ the method of AdaBoost~\cite{adaboost} to boost each QSVM in the ensemble, for a total number of
boosting iterations $G$. Boosting is a common technique in which a sequence of classifiers is trained, with 
each classifier in the sequence having enhanced focus on the training examples which were misclassified in the previous generation.
We therefore train a collection of $NG$ total QSVMs (see Figure~\ref{fig:ft_cartoon}).
The result of using this boosted ensemble scheme with the standard QSVM architecture of Figure~\ref{fig:1}(a) is 
shown in Figure~\ref{fig:results}(a)
as a function of both the number of events used to train each QSVM and the AdaBoost generation number,
for 50,000 test events.
We obtain substantial gains over using a single large QSVM, more than doubling the tagging efficiency on the test data.
In fact, even at the first generation of the AdaBoost procedure (before any boosting has taken place) we observe a considerable
improvement over the single QSVM. We attribute this gain to an interference effect in which the various QSVMs in the ensemble
``overfit in different ways'', leading on average to a stronger classifier. 
Through these improvements we achieve a peak tagging efficiency of 28.0\% on the test set, 
approaching the classical results.\\

Next we investigate the performance of the continuous variable based QSVMs depicted in Figure~\ref{fig:1}(c) and defined in Equation~\ref{eq:embed}.
As our CV data encoding scheme requires a qumode for each element of the input data vector (Equation~\ref{eq:disp}), 
and each event consists of 130 dimensional vectors~\cite{flavourtagging}, we are required to run simulations of $3\times130=390$ qubits
(recall that we simulate each qumode via three qubits~\cite{stavenger2022bosonic}).
Unfortunately, as the requirements of simulating generic circuits grows exponentially with the number of qubits,
simulating arbitrary circuits of this size is computationally intractable.
Because of this computational restriction, in order to test CV-QSVMs of this form 
using all 130 datapoints available in an event we are forced to restrict to circuits with $l=1$ (see Equation~\ref{eq:embed}).
Such circuits possess entanglement only between the (sets of three) qubits which make up a given qumode, with no 
inter-mode interactions (see Figure~\ref{fig:1}(c))
and so are easy to simulate classically. The resulting tagging efficiencies for boosted ensembles of 200 CV-QSVMs are shown 
in Figure~\ref{fig:results}(b) for varying amounts of training data. 
Although the restriction to circuits with depth $l=1$ significantly reduces their power, we find that QSVMs constructed in this way achieve results 
competitive with those reported previously via classical ML methods ($30.0\%$ via fast boosted decision trees and $28.8\%$ via deep neural networks~\cite{flavourtagging}),
reaching $29.2\%$ tagging efficiency on a test set of 50,000 events using an ensemble of 200 QSVMs each trained on 50,000 events.
A breakdown of the wrong tag and total event fractions for each bin for this classifier is given in Table~\ref{table:results}.
Due to the Gaussian nature of the displacement embedding of the event data into the qumodes (see Figure~\ref{fig:1}(d)), this 
classifier is (approximately) an ensemble of radial basis function SVMs, and is therefore essentially a classical model. 
In order to investigate the performance of deeper CV-QSVMs with nontrivial interactions between the various qumodes  
(i.e. $l>1$ in Equation~\ref{eq:embed}) we perform a PCA 
transformation on the event data to reduce its dimensionality from 130 to 5, resulting in circuits with a manageable $3\times 5=15$ qubits
and a level of complexity at which it is feasible execute deep, highly entangling feature maps.
\\

The performance of the CV-QSVMs on the PCA reduced data is shown in Figure~\ref{fig:results}(d). We find that deeper, more 
expressive feature maps significantly outperform the simple $l=1$ map on the PCA reduced data.
We also evaluate the qubit-based QSVMs on the reduced dataset, with the results shown in Figure~\ref{fig:results}(c).
As with the full 130 dimensional event data, we find that the CV models are capable of outperforming their generic 
qubit-based counterparts. 
Although the raw tagging efficiency on the five component PCA data (unsurprisingly) suffers greatly from the reduction of dimensionality, 
the relative increase in efficiency by moving to deeper circuits displayed in Figure~\ref{fig:results}(d) is encouraging,
with the highly entangled, more difficult to classically simulate CV-QSVMs recording considerable improvements in 
performance over the separable $l=1$ case.
This hints at the prospect of achieving stronger results by increasing the depth of the $l=1$ maps 
of Figure~\ref{fig:results}(b) which classify the full data.
Such circuits, however, are beyond both our ability to simulate classically
and the capabilities of the noisy, small scale quantum computers available today.

\begin{center}
  \begin{table}
    \begin{tabular}{|c|c|c|} 
 \hline
    $r$ -- interval & $w_i$ & $\epsilon_i$\\ 
 \hline\hline
      0.000\ -\ 0.100 & 0.482 & 0.159 \\
 \hline
      0.100\ -\ 0.250 & 0.400 & 0.215 \\
 \hline
      0.250\ -\ 0.500 & 0.272 & 0.292 \\
 \hline
      0.500\ -\ 0.625 & 0.158 & 0.120 \\
 \hline
     \ 0.625\ -\ 0.750 \ & 0.091 & 0.103 \\
 \hline
    \   0.750\ -\ 0.875 \ & 0.043 & 0.074 \\
 \hline
      0.875\ -\ 1.000 &\ 0.015 \ & 0.036 \\
 \hline
 %     \multicolumn{1}{ c }{} & \multicolumn{1}{ c |}{} &\  0.28\ \ \\% \cline{1-6}
 %\cline{3-3}
\end{tabular}
    \caption{The wrong tag fraction $w_i$, fraction of total events $\epsilon_i$ and $r$ -- interval of the $i$'th bin. 
    For low values of $r$ the classifier is essentially randomly guessing, with a wrong tag fraction $w_0\approx 0.5$. 
    More generally, the relationship $\expval{r_i} \approx 1-2w_i $ is observed as expected~\cite{bfactories} (see Figure~\ref{fig:bins}, Appendix A) }
    \label{table:results}
\end{table}
\end{center}

\section{Conclusion}
Machine learning has come to play an important role in the analysis of high energy physics data,
and is only expected to increase in usefulness as the amount of data created by particle accelerators increases in the future, 
as for example in the planned high luminosity upgrade to the LHC~\cite{collaborations2019report}.
The prospect of using QML methods to augment and, hopefully, improve upon these classical techniques has been
widely recognised in the particle physics community, with forward-looking studies having been already undertaken despite 
quantum computing hardware remaining in its infancy~\cite{blance2021quantum,guan2021quantum,terashi2021event,heredge2021quantum}.
In this work we have performed large scale simulations of qubit and continuous variable based QSVMs, finding that they can achieve results 
competitive with those from the classical ML algorithms which are currently employed in practice.
This is achieved despite our inability to simulate a large quantum computer in full generality, and subsequent restriction to studying 
the small class of quantum feature maps that are efficiently classically simulable, with promising results achieved despite this heavy restriction.
A full investigation of the performance of arbitrary quantum feature maps on high dimensional data such as that produced at Belle-II
must wait for the emergence of physcial quantum computers of sufficiently 
high quality to manipulate several hundred qubits (or qumodes) with high fidelity.
Excitingly, according to the published roadmaps of major quantum hardware developers~\cite{ibm_roadmap,google_roadmap,ionq_roadmap}, 
such capability may be only a few years away.\\

\noindent
\textbf{Acknowledgements:} MTW acknowledges the support of the Australian Government Research Training Program Scholarship. 
MS is supported by Australian Research Council Discovery Project DP210102831. 
Computational resources were provided by the National Computing Infrastructure (NCI) and Pawsey Supercomputing Center 
through the National Computational Merit Allocation Scheme (NCMAS).
This research was supported by The University of Melbourne’s Research Computing Services and the Petascale Campus Initiative.\\ 

The authors acknowledge the Belle-II collaboration for granting us permission to use fully simulated data for the Belle-II experiment.
The code which supports the findings of this article is available at \url{https://github.com/maxwest97/qsvm-boosted-ensemble/}.
\\ \\
\noindent
\textbf{Competing financial interests:} The authors declare no competing financial or non-financial interests.
\newpage
\def\bibsection{\subsection*{\refname}}

\bibliographystyle{naturemag}
\bibliography{./refs}

\clearpage
  \cleardoublepage
%\newpage
%\clearpage

\twocolumngrid
%\section{Appendix}
\appendix
%\section{}
\begin{center}
{\bfseries Appendix A}
\end{center}
The design of our ensembles of QSVMs entails many hyperparameter choices, including the number $N$ of QSVMs
in the ensembles, the regularisation strength $C_{\mathrm{reg}}$ and the choice of tagging bin boundaries.
In this appendix we investigate the effect of these choices.\\
\onecolumngrid

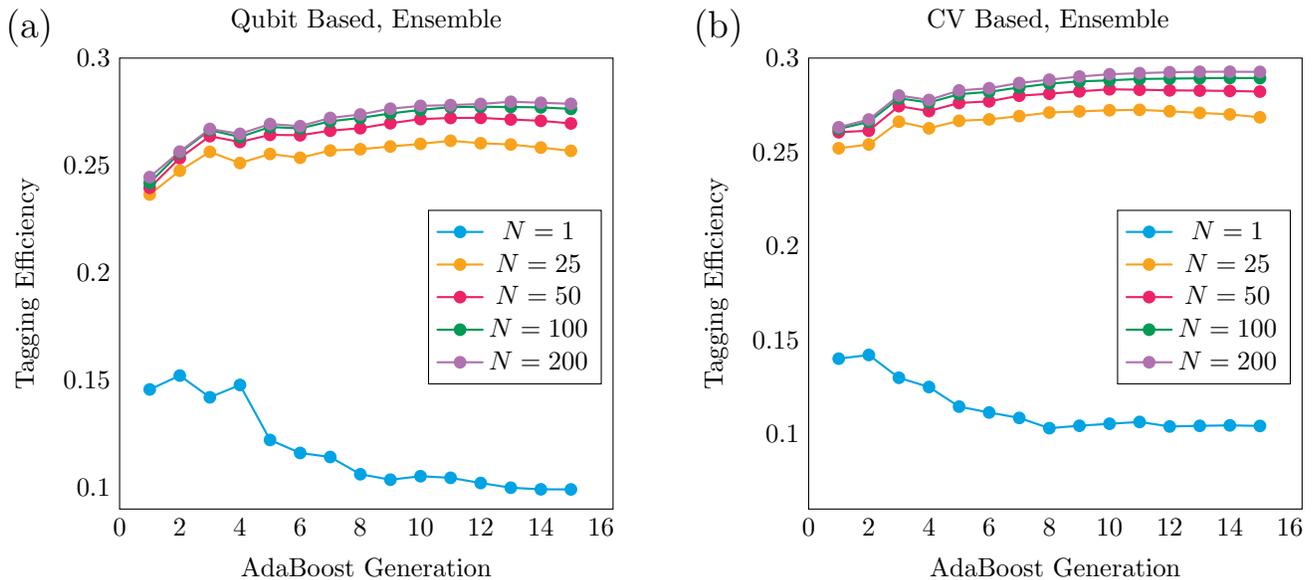
\begin{figure*}[!h]
 \begin{tikzpicture}[line cap=round]
  \node at (-1.2,6.4) {\Large (a)};
  \begin{axis}[
      domain=-1.5:3,samples=80,
      xmin=0, xmax=16.4,
      ymin=0.09, ymax=0.3,
%      %axis lines=center,
      anchor=origin,x=0.4cm,y=28.57cm, % coincide with TikZ coordinates
      ytick style={draw=none},
      xtick style={draw=none},
      title={Qubit Based, Ensemble},
      %axis equal image, % fit TikZ image
      %grid=both,
      xlabel=AdaBoost Generation,
      ylabel=Tagging Efficiency,
      %ylabel=$y$,
    legend style={at={(0.625,0.47)},anchor=west}
    ]
    %\addplot[thick,blue, scatter,scatter/classes={a={blue},b={blue},c={blue},d={blue},e={blue}}] table {testdata.txt};
    %\addplot[thick,blue, scatter,scatter/classes={a={blue}}] table {testdata.txt};
    \addplot[thick,Cerulean,skip coords between index={15}{75}, scatter,scatter/classes={a={Cerulean}}] table {circuit_vary_n.txt};
    \addplot[thick,YellowOrange,skip coords between index={0}{15},skip coords between index={30}{75}, scatter,scatter/classes={b={YellowOrange}}] table {circuit_vary_n.txt};
    \addplot[thick,WildStrawberry,skip coords between index={0}{30},skip coords between index={45}{75}, scatter,scatter/classes={c={WildStrawberry}}] table {circuit_vary_n.txt};
    \addplot[thick,ForestGreen,skip coords between index={0}{45},skip coords between index={60}{75}, scatter,scatter/classes={d={ForestGreen}}] table {circuit_vary_n.txt};
    \addplot[thick,Orchid,skip coords between index={0}{60}, scatter,scatter/classes={e={Orchid}}] table {circuit_vary_n.txt};
    \legend{$N=1$, $N=25$, $N=50$, $N=100$, $N=200$ }
  \end{axis}
\end{tikzpicture}
\qquad
 \begin{tikzpicture}[line cap=round]
  \node at (-1.2,6.4) {\Large (b)};
  \begin{axis}[
      domain=-1.5:3,samples=80,
      xmin=0, xmax=16.4,
      ymin=0.06, ymax=0.3,
      anchor=origin,x=0.4cm,y=25cm, % coincide with TikZ coordinates
      ytick style={draw=none},
      xtick style={draw=none},
      %axis equal image, % fit TikZ image
      %grid=both,
      title={CV Based, Ensemble\phantom{p}},
      xlabel=AdaBoost Generation,
      ylabel=Tagging Efficiency,
      yticklabel style={
     /pgf/number format/precision=3,
     /pgf/number format/fixed},
     legend columns=1,
      %ylabel=$y$,
    legend style={at={(0.625,0.47)},anchor=west}
    ]
    %\addplot[thick,blue, scatter,scatter/classes={a={blue},b={blue},c={blue},d={blue},e={blue}}] table {testdata.txt};
    %\addplot[thick,blue, scatter,scatter/classes={a={blue}}] table {testdata.txt};
    \addplot[thick,Cerulean,skip coords between index={15}{105}, scatter,scatter/classes={a={Cerulean}}] table {qumode_vary_n.txt};
    \addplot[thick,YellowOrange,skip coords between index={0}{15},skip coords between index={30}{105}, scatter,scatter/classes={b={YellowOrange}}] table {qumode_vary_n.txt};
    \addplot[thick,WildStrawberry,skip coords between index={0}{30},skip coords between index={45}{105}, scatter,scatter/classes={c={WildStrawberry}}] table {qumode_vary_n.txt};
    \addplot[thick,ForestGreen,skip coords between index={0}{45},skip coords between index={60}{105}, scatter,scatter/classes={d={ForestGreen}}] table {qumode_vary_n.txt};
    \addplot[thick,Orchid,skip coords between index={0}{60},skip coords between index={75}{105}, scatter,scatter/classes={e={Orchid}}] table {qumode_vary_n.txt};
    \addplot[thick,CadetBlue,skip coords between index={0}{75},skip coords between index={90}{105}, scatter,scatter/classes={f={CadetBlue}}] table {qumode_vary_n.txt};
    %\addplot[thick,Gray,skip coords between index={0}{90}, scatter,scatter/classes={g={Gray}}] table {pca_data.txt};
    \legend{$N=1$, $N=25$, $N=50$, $N=100$, $N=200$ }
    legend cell align=left,
    %\legend{$l=1$,$l=2$,$l=3$,$l=4$,$l=5$,$l=6$,RBF}
  \end{axis}
     %\node at (3.3,0.2) {\scriptsize 200 CV-QSVM Ensemble, 50000 Training Events};
     %\node at (3.865,5.5) {Top 5 PCA Components};
     %\node at (5.,0.5) {\scriptsize 50000 Training Events};
     %\node at (4.825,0.2) {\scriptsize 200 CV-QSVM Ensemble};
\end{tikzpicture}
\caption{\textbf{Ensemble size}. We obtain greatly increased performance by considering ensemble classifiers 
instead of an individual QSVM.
In both the qubit and CV cases we find that there is little benefit in increasing the size 
$N$ of the ensemble beyond $N\gtrsim 100$. }
\label{fig:esize}
\end{figure*}
%\phantom{...}
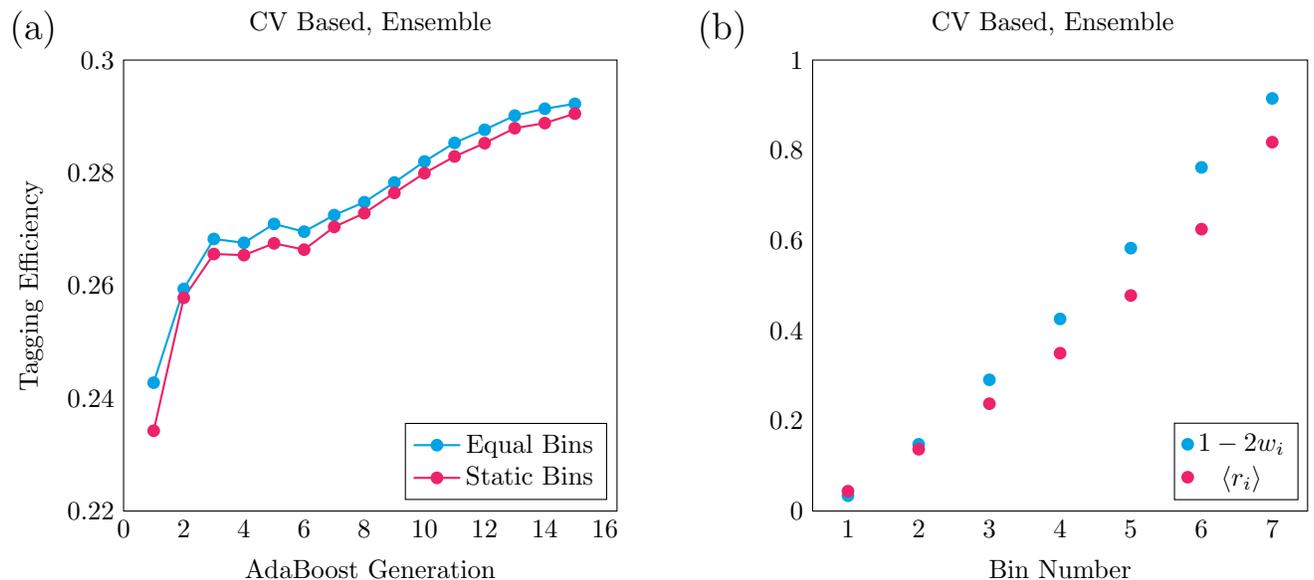
\begin{figure*}[!h]
 \begin{tikzpicture}[line cap=round]
  \node at (-1.2,6.4) {\Large (a)};
  \begin{axis}[
      domain=-1.5:3,samples=80,
      xmin=0, xmax=16.4,
      ymin=0.22, ymax=0.3,
      anchor=origin,x=0.4cm,y=75cm, % coincide with TikZ coordinates
      ytick style={draw=none},
      xtick style={draw=none},
      title={CV Based, Ensemble},
      %axis equal image, % fit TikZ image
      %grid=both,
      xlabel=AdaBoost Generation,
      ylabel=Tagging Efficiency,
      %ylabel=$y$,
    legend style={at={(0.57,0.11)},anchor=west}
    ]
    %\addplot[thick,blue, scatter,scatter/classes={a={blue},b={blue},c={blue},d={blue},e={blue}}] table {testdata.txt};
    %\addplot[thick,blue, scatter,scatter/classes={a={blue}}] table {testdata.txt};
    \addplot[thick,Cerulean,skip coords between index={15}{30}, scatter,scatter/classes={a={Cerulean}}] table {bin_comp.txt};
    \addplot[thick,WildStrawberry,skip coords between index={0}{15}, scatter,scatter/classes={b={WildStrawberry}}] table {bin_comp.txt};
    \legend{Equal Bins, Static Bins }
  \end{axis}
\end{tikzpicture}
\qquad
 \begin{tikzpicture}[line cap=round]
  \node at (-1.2,6.4) {\Large (b)};
  \begin{axis}[
      domain=-1.5:3,samples=80,
      xmin=0.5, xmax=7.5,
      ymin=0., ymax=1,
      anchor=origin,x=0.94cm,y=6.cm, % coincide with TikZ coordinates
      ytick style={draw=none},
      xtick style={draw=none},
      %axis equal image, % fit TikZ image
      %grid=both,
      title={CV Based, Ensemble\phantom{p}},
      xlabel=Bin Number,
      yticklabel style={
     /pgf/number format/precision=3,
     /pgf/number format/fixed},
     legend columns=1,
      %ylabel=$y$,
    legend style={at={(0.73,0.11)},anchor=west}
    ]
    %\addplot[thick,blue, scatter,scatter/classes={a={blue},b={blue},c={blue},d={blue},e={blue}}] table {testdata.txt};
    %\addplot[thick,blue, scatter,scatter/classes={a={blue}}] table {testdata.txt};
    \addplot[thick,only marks, Cerulean,skip coords between index={7}{14}, scatter,scatter/classes={a={Cerulean}}] table {bin_vary.txt};
    \addplot[thick,only marks, WildStrawberry,skip coords between index={0}{7}, scatter,scatter/classes={b={WildStrawberry}}] table {bin_vary.txt};
    %\addplot[thick,Gray,skip coords between index={0}{90}, scatter,scatter/classes={g={Gray}}] table {pca_data.txt};
    \legend{ $1-2w_i$, $\expval{r_i}$}
    legend cell align=left,
    %\legend{$l=1$,$l=2$,$l=3$,$l=4$,$l=5$,$l=6$,RBF}
  \end{axis}
     %\node at (3.3,0.2) {\scriptsize 200 CV-QSVM Ensemble, 50000 Training Events};
     %\node at (3.865,5.5) {Top 5 PCA Components};
     %\node at (5.,0.5) {\scriptsize 50000 Training Events};
     %\node at (4.825,0.2) {\scriptsize 200 CV-QSVM Ensemble};
\end{tikzpicture}
\caption{\textbf{Bin statistics.} (a) We consider both the standard Belle-II strategy of employing bins 
with ranges $[0,0.1],[0.1,0.25],[0.25,0.5],[0.5,0.625],[0.625,0.75],[0.75,0.875],[0.875,1.0]$ (see Table~\ref{table:results}),
and with bins whose width is dynamically adjusted so that the number of events in each bin is equal. 
We find the benefit of doing this is minimal, however.
(b) For each bin we observe $1-2w_i\approx \expval{r_i}$, as expected. 
The results of this Figure are calculated using an ensemble of 200 $l=1$ CV-QSVMs as 
in Figure~\ref{fig:results}(b) with each CV-QSVM trained on 50,000 events. }
\label{fig:bins}
\end{figure*}

\begin{figure*}[!h]
 \begin{tikzpicture}[line cap=round]
  \begin{axis}[
      domain=-1.5:3,samples=80,
      xmin=10, xmax=110,
      ymin=0., ymax=1.03,
      anchor=origin,x=0.0305cm,y=5.25cm, % coincide with TikZ coordinates
      ytick style={draw=none},
      xtick style={draw=none},
      title={$C_{\mathrm{reg}} = 0.1$},
      %axis equal image, % fit TikZ image
      %grid=both,
      xlabel style={align=center},xlabel=Training Events\\ $(\times 10^3)$,
      ylabel=Tagging Efficiency,
      %ylabel=$y$,
    legend style={at={(0.36,0.25)},anchor=west}
    ]
    %\addplot[thick,blue, scatter,scatter/classes={a={blue},b={blue},c={blue},d={blue},e={blue}}] table {testdata.txt};
    %\addplot[thick,blue, scatter,scatter/classes={a={blue}}] table {testdata.txt};
    \addplot[thick,Cerulean,skip coords between index={5}{10}, scatter,scatter/classes={a={Cerulean}}] table {reg_0.1};
    \addplot[thick,WildStrawberry,skip coords between index={0}{5}, scatter,scatter/classes={b={WildStrawberry}}] table {reg_0.1};
    \legend{Train, Test}
  \end{axis}
\end{tikzpicture}
 \begin{tikzpicture}[line cap=round]
  \begin{axis}[
      domain=-1.5:3,samples=80,
      xmin=10, xmax=110,
      ymin=0., ymax=1.03,
      anchor=origin,x=0.0305cm,y=5.25cm, % coincide with TikZ coordinates
      ytick style={draw=none},
      xtick style={draw=none},
      yticklabels={},
      title={$C_{\mathrm{reg}} = 0.5$},
      %axis equal image, % fit TikZ image
      %grid=both,
      xlabel style={align=center},xlabel=Training Events\\ $(\times 10^3)$,
      %ylabel=Tagging Efficiency,
      %ylabel=$y$,
    legend style={at={(0.36,0.25)},anchor=west}
    ]
    %\addplot[thick,blue, scatter,scatter/classes={a={blue},b={blue},c={blue},d={blue},e={blue}}] table {testdata.txt};
    %\addplot[thick,blue, scatter,scatter/classes={a={blue}}] table {testdata.txt};
    \addplot[thick,Cerulean,skip coords between index={5}{10}, scatter,scatter/classes={a={Cerulean}}] table {reg_0.5};
    \addplot[thick,WildStrawberry,skip coords between index={0}{5}, scatter,scatter/classes={b={WildStrawberry}}] table {reg_0.5};
    \legend{Train, Test}
  \end{axis}
\end{tikzpicture}
 \begin{tikzpicture}[line cap=round]
  \begin{axis}[
      domain=-1.5:3,samples=80,
      xmin=10, xmax=110,
      ymin=0., ymax=1.03,
      anchor=origin,x=0.0305cm,y=5.25cm, % coincide with TikZ coordinates
      ytick style={draw=none},
      xtick style={draw=none},
      yticklabels={},
      title={$C_{\mathrm{reg}} = 1.0$},
      %axis equal image, % fit TikZ image
      %grid=both,
      xlabel style={align=center},xlabel=Training Events\\ $(\times 10^3)$,
      %ylabel=Tagging Efficiency,
      %ylabel=$y$,
    legend style={at={(0.36,0.25)},anchor=west}
    ]
    %\addplot[thick,blue, scatter,scatter/classes={a={blue},b={blue},c={blue},d={blue},e={blue}}] table {testdata.txt};
    %\addplot[thick,blue, scatter,scatter/classes={a={blue}}] table {testdata.txt};
    \addplot[thick,Cerulean,skip coords between index={5}{10}, scatter,scatter/classes={a={Cerulean}}] table {reg_1.0};
    \addplot[thick,WildStrawberry,skip coords between index={0}{5}, scatter,scatter/classes={b={WildStrawberry}}] table {reg_1.0};
    \legend{Train, Test}
  \end{axis}
\end{tikzpicture}
 \begin{tikzpicture}[line cap=round]
  \begin{axis}[
      domain=-1.5:3,samples=80,
      xmin=10, xmax=110,
      ymin=0., ymax=1.03,
      anchor=origin,x=0.0305cm,y=5.25cm, % coincide with TikZ coordinates
      ytick style={draw=none},
      xtick style={draw=none},
      yticklabels={},
      title={$C_{\mathrm{reg}} = 5.0$},
      %axis equal image, % fit TikZ image
      %grid=both,
      xlabel style={align=center},xlabel=Training Events\\ $(\times 10^3)$,
      %ylabel=Tagging Efficiency,
      %ylabel=$y$,
    legend style={at={(0.36,0.25)},anchor=west}
    ]
    %\addplot[thick,blue, scatter,scatter/classes={a={blue},b={blue},c={blue},d={blue},e={blue}}] table {testdata.txt};
    %\addplot[thick,blue, scatter,scatter/classes={a={blue}}] table {testdata.txt};
    \addplot[thick,Cerulean,skip coords between index={5}{10}, scatter,scatter/classes={a={Cerulean}}] table {reg_5.0};
    \addplot[thick,WildStrawberry,skip coords between index={0}{5}, scatter,scatter/classes={b={WildStrawberry}}] table {reg_5.0};
    \legend{Train, Test}
  \end{axis}
\end{tikzpicture}
 \begin{tikzpicture}[line cap=round]
  \begin{axis}[
      domain=-1.5:3,samples=80,
      xmin=10, xmax=110,
      ymin=0., ymax=1.03,
      anchor=origin,x=0.0305cm,y=5.25cm, % coincide with TikZ coordinates
      ytick style={draw=none},
      xtick style={draw=none},
      yticklabels={},
      title={$C_{\mathrm{reg}} = 10.0$},
      %axis equal image, % fit TikZ image
      %grid=both,
      xlabel style={align=center},xlabel=Training Events\\ $(\times 10^3)$,
      %ylabel=Tagging Efficiency,
      %ylabel=$y$,
    legend style={at={(0.36,0.25)},anchor=west}
    ]
    %\addplot[thick,blue, scatter,scatter/classes={a={blue},b={blue},c={blue},d={blue},e={blue}}] table {testdata.txt};
    %\addplot[thick,blue, scatter,scatter/classes={a={blue}}] table {testdata.txt};
    \addplot[thick,Cerulean,skip coords between index={5}{10}, scatter,scatter/classes={a={Cerulean}}] table {reg_10.0};
    \addplot[thick,WildStrawberry,skip coords between index={0}{5}, scatter,scatter/classes={b={WildStrawberry}}] table {reg_10.0};
    \legend{Train, Test}
  \end{axis}
\end{tikzpicture}
\caption{\textbf{Regularisation.} Although modifying the constant $C_{\mathrm{reg}}$ which controls the strength of the 
regularisation~\cite{scikit-learn} can reduce the overfitting of a single qubit-based QSVM to the training data, we find 
that it is incapable of significantly increasing the performance on the test data, motivating our consideration of ensemble classifiers.
The QSVM used here is as described in Figure~\ref{fig:overfit}.
}
\label{fig:reg}
\end{figure*}
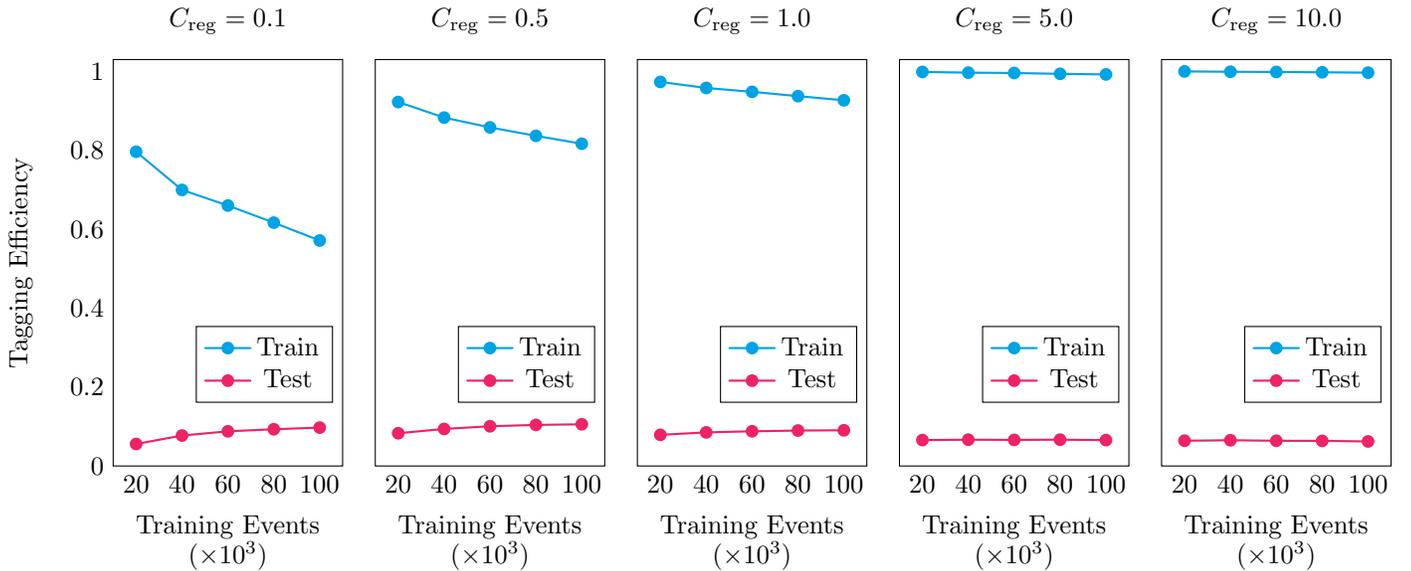
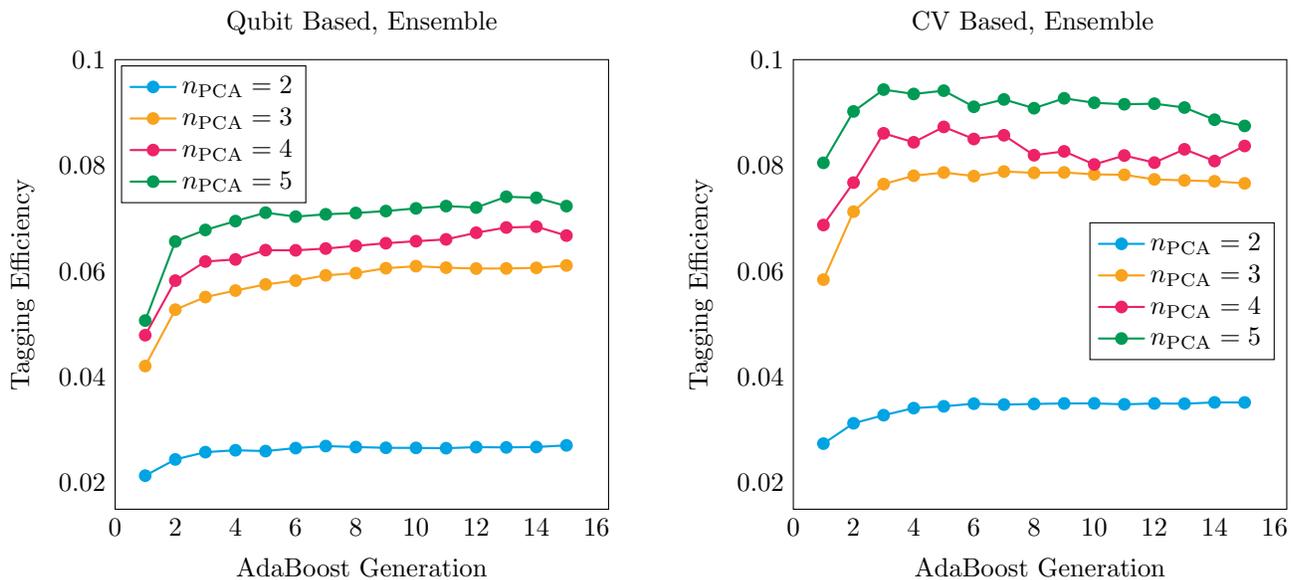
\begin{figure*}[!h]
 \begin{tikzpicture}[line cap=round]
  %\node at (-1.2,6.4) {\Large (a)};
  \begin{axis}[
      domain=-1.5:3,samples=80,
      xmin=0, xmax=16.4,
      ymin=0.015, ymax=0.1,
%      %axis lines=center,
      anchor=origin,x=0.4cm,y=70.37cm, % coincide with TikZ coordinates
      ytick style={draw=none},
      xtick style={draw=none},
      title={Qubit Based, Ensemble},
      %axis equal image, % fit TikZ image
      %grid=both,
      xlabel=AdaBoost Generation,
      ylabel=Tagging Efficiency,
      yticklabel style={
     /pgf/number format/precision=3,
     /pgf/number format/fixed},
      %ylabel=$y$,
    legend style={at={(0.012,0.834)},anchor=west}
    ]
    %\addplot[thick,blue, scatter,scatter/classes={a={blue},b={blue},c={blue},d={blue},e={blue}}] table {testdata.txt};
    %\addplot[thick,blue, scatter,scatter/classes={a={blue}}] table {testdata.txt};
    \addplot[thick,Cerulean,skip coords between index={15}{60}, scatter,scatter/classes={a={Cerulean}}] table {circuit_vary_pca.txt};
    \addplot[thick,YellowOrange,skip coords between index={0}{15},skip coords between index={30}{60}, scatter,scatter/classes={b={YellowOrange}}] table {circuit_vary_pca.txt};
    \addplot[thick,WildStrawberry,skip coords between index={0}{30},skip coords between index={45}{60}, scatter,scatter/classes={c={WildStrawberry}}] table {circuit_vary_pca.txt};
    \addplot[thick,ForestGreen,skip coords between index={0}{45},scatter,scatter/classes={d={ForestGreen}}] table {circuit_vary_pca.txt};
    \legend{$n_{\mathrm{PCA}}=2$,$n_{\mathrm{PCA}}=3$,$n_{\mathrm{PCA}}=4$,$n_{\mathrm{PCA}}=5$ }
  \end{axis}
\end{tikzpicture}
\qquad
 \begin{tikzpicture}[line cap=round]
  %\node at (-1.2,6.4) {\Large (b)};
  \begin{axis}[
      domain=-1.5:3,samples=80,
      xmin=0, xmax=16.4,
      ymin=0.015, ymax=0.1,
      anchor=origin,x=0.4cm,y=70.37cm, % coincide with TikZ coordinates
      ytick style={draw=none},
      xtick style={draw=none},
      %axis equal image, % fit TikZ image
      %grid=both,
      title={CV Based, Ensemble\phantom{p}},
      xlabel=AdaBoost Generation,
      ylabel=Tagging Efficiency,
      yticklabel style={
     /pgf/number format/precision=3,
     /pgf/number format/fixed},
     legend columns=1,
      %ylabel=$y$,
    legend style={at={(0.6,0.485)},anchor=west}
    ]
    %\addplot[thick,blue, scatter,scatter/classes={a={blue},b={blue},c={blue},d={blue},e={blue}}] table {testdata.txt};
    %\addplot[thick,blue, scatter,scatter/classes={a={blue}}] table {testdata.txt};
    \addplot[thick,Cerulean,skip coords between index={15}{60}, scatter,scatter/classes={a={Cerulean}}] table {qumode_vary_pca.txt};
    \addplot[thick,YellowOrange,skip coords between index={0}{15},skip coords between index={30}{60}, scatter,scatter/classes={b={YellowOrange}}] table {qumode_vary_pca.txt};
    \addplot[thick,WildStrawberry,skip coords between index={0}{30},skip coords between index={45}{60}, scatter,scatter/classes={c={WildStrawberry}}] table {qumode_vary_pca.txt};
    \addplot[thick,ForestGreen,skip coords between index={0}{45}, scatter,scatter/classes={d={ForestGreen}}] table {qumode_vary_pca.txt};
    %\addplot[thick,Gray,skip coords between index={0}{90}, scatter,scatter/classes={g={Gray}}] table {pca_data.txt};
    \legend{$n_{\mathrm{PCA}}=2$,$n_{\mathrm{PCA}}=3$,$n_{\mathrm{PCA}}=4$,$n_{\mathrm{PCA}}=5$ }
    legend cell align=left,
    %\legend{$l=1$,$l=2$,$l=3$,$l=4$,$l=5$,$l=6$,RBF}
  \end{axis}
     %\node at (3.3,0.2) {\scriptsize 200 CV-QSVM Ensemble, 50000 Training Events};
     %\node at (3.865,5.5) {Top 5 PCA Components};
     %\node at (5.,0.5) {\scriptsize 50000 Training Events};
     %\node at (4.825,0.2) {\scriptsize 200 CV-QSVM Ensemble};
\end{tikzpicture}
\caption{\textbf{Number of PCA Components}. We find steady improvements in peak tagging efficiency 
when making more PCA components (out of a total of 130) available to the QSVM ensembles.
With each qumode in the CV-QSVMs being simulated with three qubits, we are able to simulate CV-QSVMs
accepting up to five PCA components, corresponding to 15 qubits in the backend implementation. 
The architectures of the QSVMs employed in this Figure are the same as 
the highest performing architectures in Figure~\ref{fig:results}(c,d) (i.e. $d=50$ in the qubit case and $l=6$
in the CV case).
}
\label{fig:pca_comps}
\end{figure*}
\end{document}